\documentclass[lettersize,journal]{IEEEtran}
\usepackage{amsmath,amsfonts}
\usepackage{algorithm}
\usepackage{algpseudocode}
\usepackage{array}
\usepackage{textcomp}
\usepackage{stfloats}
\usepackage{url}
\usepackage{verbatim}
\usepackage{graphicx}
\usepackage{cite}
\usepackage{indentfirst}
\usepackage{multirow}
\usepackage{color}
\usepackage{xcolor}
\usepackage{colortbl}
\usepackage{listings}
\usepackage{enumerate}
\usepackage{amsthm,amscd,amssymb,latexsym,upref,stmaryrd}
\usepackage{epsfig}
\usepackage{float}
\usepackage{lipsum}
\usepackage{multicol}
\usepackage{setspace}
\usepackage{bm}
\usepackage{arydshln}
\usepackage{booktabs}
\usepackage{threeparttable}
\usepackage{stackengine}
\usepackage[utf8]{inputenc}
\usepackage{fancyhdr}
\usepackage[colorlinks,linkcolor=blue,anchorcolor=blue,citecolor=blue]{hyperref}
\usepackage{makecell}
\usepackage{subcaption}

\newtheorem{remark}{Remark}
\newtheorem{theorem}{Theorem}

\newcommand{\calosymbol}[1]{\text{\usefont{U}{BOONDOX-calo}{m}{n}#1}}
\newcommand{\xrowht}[2][0]{\addstackgap[.5\dimexpr#2\relax]{\vphantom{#1}}}

\begin{document}

\title{RIS-aided Single-frequency 3D Imaging \\ by Exploiting Multi-view Image Correlations}

\author{Yixuan~Huang,~\IEEEmembership{Graduate Student Member,~IEEE,} Jie~Yang,~\IEEEmembership{Member,~IEEE,} \\ \  Chao-Kai~Wen,~\IEEEmembership{Fellow,~IEEE,} and Shi~Jin,~\IEEEmembership{Fellow,~IEEE}
\thanks{This work was supported in part by the National Key Research and Development Program of China 2018YFA0701602;
in part by the National Natural Science Foundation of China (NSFC) under Grant 62261160576, Grant 62301156, and Grant 62341107;
in part by the Key Technologies R\&D Program of Jiangsu (Prospective and Key Technologies for Industry) under Grant BE2023022 and Grant BE2023022-1;
in part by the Fundamental Research Funds for the Central Universities 2242022k60004;
in part by the National Science Foundation of Jiangsu Province under Grant BK20230818.
The work of C.-K. Wen was partly supported by Qualcomm via a Taiwan University Research Collaboration Project and the Sixth Generation Communication and Sensing Research Center, funded by Taiwan's MOE under the Higher Education SPROUT Project.
\emph{(Corresponding authors: Shi Jin; Jie Yang.)}

Yixuan~Huang is with the National Mobile Communications Research Laboratory, Southeast University, Nanjing 210096, China (e-mail: huangyx@seu.edu.cn).

Jie~Yang is with the Key Laboratory of Measurement and Control of Complex Systems of Engineering, Ministry of Education, and the Frontiers Science Center for Mobile Information Communication and Security, Southeast University, Nanjing 210096, China (e-mail: yangjie@seu.edu.cn).

Chao-Kai~Wen is with the Institute of Communications Engineering, National Sun Yat-sen University, Kaohsiung 80424, Taiwan. (e-mail: chaokai.wen@mail.nsysu.edu.tw).

Shi~Jin is with the National Mobile Communications Research Laboratory and the Frontiers Science Center for Mobile Information Communication and Security, Southeast University, Nanjing 210096, China (e-mail: jinshi@seu.edu.cn).

}
}

\maketitle

\begin{abstract}
Retrieving range information in three-dimensional (3D) radio imaging is particularly challenging due to the limited communication bandwidth and pilot resources.
To address this issue, we consider a reconfigurable intelligent surface (RIS)-aided uplink communication scenario, generating multiple measurements through RIS phase adjustment. This study successfully realizes 3D single-frequency imaging by exploiting the near-field multi-view image correlations deduced from user mobility.
We first highlight the significance of considering anisotropy in multi-view image formation by investigating radar cross-section properties and diffraction resolution limits. We then propose a novel model for joint multi-view 3D imaging that incorporates occlusion effects and anisotropic scattering. These factors lead to slow image support variation and smooth coefficient evolution, which are mathematically modeled as Markov processes. Based on this model, we employ the Expectation Maximization-Turbo-Generalized Approximate Message Passing algorithm for joint multi-view single-frequency 3D imaging with limited measurements.
Simulation results reveal the superiority of joint multi-view imaging in terms of enhanced imaging ranges, accuracies, and anisotropy characterization compared to single-view imaging. Combining adjacent observations for joint multi-view imaging enables a reduction in the measurement overhead by 80\%.
\end{abstract}


\section{Introduction}

Radio imaging, a critical component of environmental sensing, facilitates various applications such as monitoring, augmented reality, and environmental reconstruction, ultimately enhancing communication throughput \cite{liu2022integrated,yang2021enabling,imani2020review}. Its advantages include irrelevance to lighting conditions and protection of human privacy. 
By integrating imaging within communication systems, ubiquitous fine-grained sensing without extra infrastructure is anticipated \cite{wei2022toward}.
In this context, the primary technological focus is on retrieving the scattering coefficient distribution of the region of interest (ROI), which is discretized into numerous voxels. The scattering coefficient, or root radar cross-section (RCS) of a voxel, is the root of the energy it scatters when illuminated by a unit energy field. While synthetic aperture radar (SAR) imaging techniques \cite{sheen2001three} have been widely studied and employed in radio imaging, they cannot be directly applied to communication systems due to small antenna apertures, limited bandwidths, low pilot overheads, and waveform mismatches \cite{liu2022integrated,lien20175g}.

Recently, metasurfaces have inspired numerous innovative designs for radio imaging, where electromagnetic (EM) environments are manually customized to capture target scattering characteristics \cite{imani2020review}.
As a type of metasurfaces, reconfigurable intelligent surfaces (RISs), comprising an array of subwavelength tunable elements, are considered essential for future communication systems, assisting in communication and environment sensing with low energy consumption \cite{tang2021path,rahal2021ris,chen2023multi}. However, RIS-aided imaging is more challenging than traditional SAR imaging, as RIS elements can only reflect signals but cannot receive or transmit them \cite{tang2021path}. Leveraging the rapid reconfiguration capabilities \cite{castaldi2021joint}, RIS-aided imaging based on beamforming and scanning is realized in \cite{taha2022reconfigurable}, whereas another strategy of specially designed phase shifts is studied in \cite{jiang2023near}, both requiring a large number of measurements.
Considering the sparse nature of the targets in the ROI \cite{ccetin2014sparsity}, compressed sensing (CS)-based algorithms, such as subspace pursuit (SP) \cite{dai2009subspace} and generalized approximate message passing (GAMP) \cite{rangan2011generalized}, can be employed to reconstruct ROI images with RISs \cite{hu2022metasketch,sankar2023coded}, requiring fewer measurements than the Nyquist criterion \cite{ccetin2014sparsity}.
Moreover, joint communication and imaging are studied in \cite{zhu2023ris}.
These algorithms utilize channel state information obtained through channel estimation with pilot signals to perceive the scattering properties of the ROI.

However, previous studies \cite{jiang2023near,hu2022metasketch,zhu2023ris,sankar2023coded} only considered two-dimensional (2D) targets, and the imaging capabilities of RIS-assisted systems have not been theoretically investigated. In this study, we extend prior work to RIS-aided three-dimensional (3D) imaging with a single frequency. Specifically, the RIS array is used as the imaging aperture, and its phase shifts can be optimized to improve imaging performances.
Moreover, we adopt the concept of point spread function (PSF) in SAR imaging \cite{gao2018efficient} and CS theory \cite{lustig2007sparse}, which refers to the response of the imaging system to a point target, to reveal the geometric constraints of subpath correlation from the ROI to the RIS on the system imaging capabilities.
Furthermore, we address the tradeoff between imaging resolutions, accuracies, and the RIS-subtended angle.

Performing 3D imaging with a single frequency is challenging, as range information retrieval heavily relies on large bandwidths \cite{sheen2001three}.
However, recent studies have demonstrated the possibility of single-frequency 3D imaging in near-field regions with multiple-input multiple-output (MIMO) antenna arrays \cite{fromenteze2017single,imani2020review}.
Nevertheless, this requires large apertures and short imaging distances, as diffraction effects impose limitations on image resolutions \cite{sheen2001three,jiang2023near}.
In communication systems, large apertures are proposed to be synthesized using distributed antennas of multiple user equipments (UEs) \cite{tong2021joint}, and multi-view imaging has been studied in \cite{tong2022environment} to address occlusion effects in 3D space, enabling image constitution with a single frequency.
However, anisotropic scattering, which has been overlooked in conventional single-view imaging due to relatively small apertures, should be considered when the target is sensed from widely distributed UE positions \cite{ccetin2014sparsity}.
Anisotropy is related to RCS properties, indicating that voxel scattering coefficients vary with observation angles, and the perceived images differ at distinct UE positions, as studied in wide-angle SAR imaging fields \cite{moses2004wide}.
In this study, we analyze the properties of RCS with respect to incident and scattering angles \cite{huang2023joint,deban2009deterministic} and combine them with diffraction resolution limits \cite{sheen2001three} to emphasize the necessity of considering anisotropic scattering properties in multi-view imaging scenarios.

\begin{table*}[t]
  \renewcommand{\arraystretch}{1.5}
  \centering
  \fontsize{8}{8}\selectfont
  \captionsetup{font=small}
  \caption{Comparison of related works.}\label{tab-paper-comp}
  \begin{threeparttable}
    \begin{tabular}{|c|p{0.45cm}<{\centering}|p{0.45cm}<{\centering}|c|c|c|c|c|c|c|}
      \hline
      \multirow{2}{*}{References} & \multicolumn{2}{c|}{Dimensions} & \multirow{2}{*}{\makecell[c]{Single \\ frequency}} & \multirow{2}{*}{\makecell[c]{RIS-aided}} & \multirow{2}{*}{\makecell[c]{Communication \\ systems}} & \multirow{2}{*}{\makecell[c]{Multi-view \\ imaging}} & \multirow{2}{*}{\makecell[c]{Occlusion \\ effects}} & \multirow{2}{*}{\makecell[c]{Anisotropic \\ scattering}} & \multirow{2}{*}{Other assistance} \\
      \cline{2-3}
      & 2D & 3D & & & & & & & \\
      \hline
      \cite{sheen2001three} & \checkmark & \checkmark & & &  &  &  & & Synthesized apertures \\
      \hline
      \cite{jiang2023near,hu2022metasketch,sankar2023coded} & \checkmark & & \checkmark & \checkmark & & &  &  &  \\
      \hline
      \cite{zhu2023ris} & \checkmark & & \checkmark & \checkmark & \checkmark &  &  &  &  \\
      \hline
      \cite{fromenteze2017single} & & \checkmark & \checkmark & &  &  &  & & MIMO array \\
      \hline
      \cite{tong2021joint} & & \checkmark & \checkmark & \checkmark & \checkmark &  &  & & Multiple UEs \\
      \hline
      \cite{tong2022environment} & & \checkmark & \checkmark &  & \checkmark & \checkmark & \checkmark & & Multiple UEs  \\
      \hline
      \cite{moses2004wide,wu2021through} & \checkmark & \checkmark &  &  &  & \checkmark &  & \checkmark &  \\
      \hline
      Our work & & \checkmark & \checkmark & \checkmark & \checkmark & \checkmark & \checkmark & \checkmark & \\
      \hline
    \end{tabular}
  \end{threeparttable}
\end{table*}

Despite the presence of occlusion effects and anisotropic scattering when uniting observations from distinct directions, multi-view imaging has been studied in conventional scenarios.
The large aperture synthesized by distributed UEs can be divided into sub-apertures where the isotropic assumption holds \cite{moses2004wide}.
However, determining proper sub-aperture sizes is challenging, as it involves a trade-off between problem ill-conditioning and model mismatch \cite{stojanovic2008joint}.
Alternatively, certain correlations exist in the sequential images observed at adjacent UE positions, which can be exploited to enhance imaging accuracy \cite{ccetin2014sparsity,ziniel2013dynamic}.
In \cite{tong2022environment}, multi-view occlusion effects are modeled as a multiplicative Bernoulli perturbation of the sensing matrix, but isotropic scattering is assumed, resulting in a single ROI image.
Instead, anisotropy has been employed to formulate dynamic signal estimation problems in \cite{wu2021through}, but a joint sparse model that assumes the same image support is used.
In this study, we introduce multi-view imaging to RIS-aided communication systems, where the UE moves along a continuous trajectory.
Distinct from previous studies that synthesize apertures with distributed UE antennas \cite{tong2021joint,tong2022environment}, occlusion effects and anisotropy are simultaneously considered to model multi-view image correlations, which are captured by the Expectation Maximization-turbo-GAMP (EM-turbo-GAMP) algorithm \cite{ziniel2012generalized}. Multi-view observation of the ROI helps range and cross-range information retrieval to achieve single-frequency 3D imaging with low pilot overhead. 

The comparison of related works is enumerated in Table \ref{tab-paper-comp}, and our key contributions are summarized as follows:

\begin{itemize}
\item \textbf{Consideration of anisotropy in multi-view imaging:}
We conduct a study on RCS properties in relation to observation angles, present the diffraction resolution limits, and establish the necessity of considering anisotropic properties in multi-view imaging.

\item \textbf{Proposal of a joint multi-view single-frequency 3D imaging scheme:}
We formulate occlusion effects and anisotropic scattering as Markov processes, propose a joint multi-view 3D imaging model with a single frequency, adopt the EM-turbo-GAMP algorithm to exploit sequential image correlations, and prove the advantages of joint multi-view imaging through simulations.

\item \textbf{Theoretical analysis of imaging performance:}
We optimize RIS phase shifts, utilize the PSF to analyze imaging capabilities, establish the geometric constraint of subpath correlation, and explore the tradeoff between imaging accuracy, resolutions, and the RIS-subtended angle.
\end{itemize}

{\bf Notations}---The scalars (e.g., $a$) are denoted in italic, vectors (e.g., $\mathbf{a}$) in bold, and matrices (e.g., $\mathbf{A}$) in bold capital letters.
Random variables of realizations $a$ and $\mathbf{a}$ are denoted as $\mathcal{A}$ and $\boldsymbol{\mathcal{A}}$, respectively.
Random process is given by $\{\mathbf{a}_t\}_{t=1}^T$.
The $\ell_{2}$-norm of $\mathbf{a}$ is $\|\mathbf{a}\|_2$, and the Frobenius norm of $\mathbf{A}$ is $\|\mathbf{A}\|_{\rm{F}}$.
$\left<\mathbf{a}, \mathbf{b}\right>$ calculates the inner product of $\mathbf{a}$ and $\mathbf{b}$, whereas $|\cdot|$ denotes the module of a complex value.
$\operatorname{supp}(\mathbf{a})$ derives the support set of $\mathbf{a}$, and $\operatorname{card}(\cdot)$ calculates the cardinality of a set.
$j = \sqrt{-1}$ is the imaginary unit.
$\mathbb{E}\{\cdot\}$ calculates statistical expectation, and $\propto$ denotes positive correlation.
Conjugate, transpose, and Hermitian operators are presented as $(\cdot)^*$, $(\cdot)^{\rm{T}}$, and $(\cdot)^{\rm{H}}$, respectively.
$\text{diag}\{\cdot\}$ denotes a diagonal matrix, and the Hadamard product is presented by $\odot$.

\section{System Model}
\label{sec:system-model}

\begin{figure}
\centering
\captionsetup{font=footnotesize}
\begin{subfigure}[b]{\linewidth}
\centering
\includegraphics[width=0.8\linewidth]{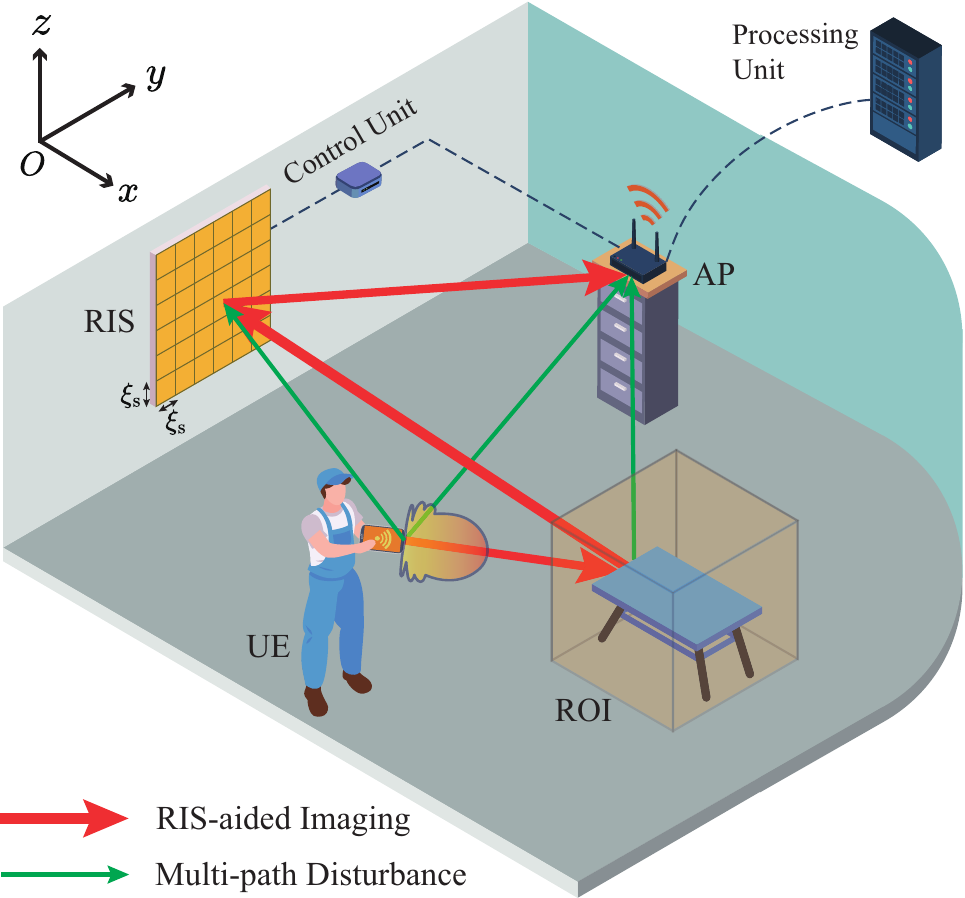}
\caption{RIS-aided communication system.}
\label{fig:model-a}
\end{subfigure}
\begin{subfigure}[b]{\linewidth}
\centering
\vspace{0.2cm}
\includegraphics[width=0.6\linewidth]{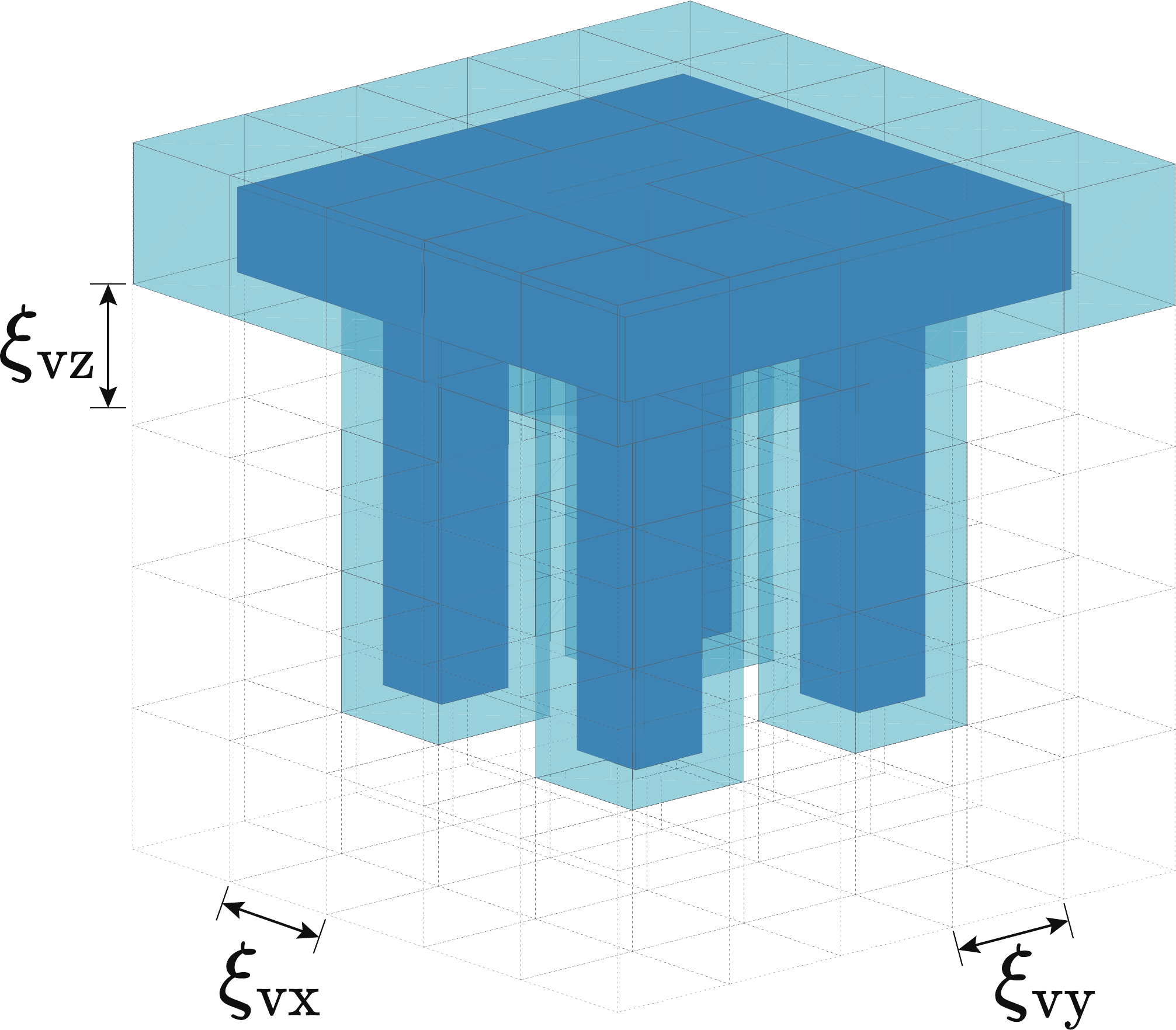}
\caption{Discretized ROI with a sparse target.}
\label{fig:model-b}
\end{subfigure}
\caption{Illustration of the considered RIS-aided communication system.}
\label{fig:model}
\end{figure}

We consider a RIS-aided communication system operating in the 3D space $\mathbb{R}^3=\{[x,y,z]^{\rm{T}}:x,y,z,\in\mathbb{R}\}$, as depicted in Fig. \ref{fig:model-a}.
The system consists of a single-antenna UE, a target in the ROI, an access point (AP) with $Q$ antennas, and a vertically placed RIS utilizing a uniform planar array (UPA) with $M$ elements.
The antenna spacings of the AP and the RIS are $\xi_{\rm{a}}$ and $\xi_{\rm{s}}$, respectively.
RIS phase shifts are controlled by the AP through a control unit, and the received signals of the AP are managed in the processing unit.
The ROI is discretized into $N = N_{\rm{x}}\times N_{\rm{y}}\times  N_{\rm{z}}$ voxels, with each voxel having a size of $\xi_{\rm{vx}} \times \xi_{\rm{vy}} \times \xi_{\rm{vz}}$, as illustrated in Fig. \ref{fig:model-b}.
In this study, the imaging resolution exactly corresponds to the voxel size, where a higher resolution is associated with a smaller voxel size.

We assume that the RIS is in the near field of the ROI\footnote{
The near field means that the distance $\calosymbol{d}$ between the ROI and the RIS should be smaller than $2r^2/\lambda$, where $r$ is the largest dimension of the ROI \cite{tang2021path}.
For example, in the simulation settings of Sec. \ref{subsubsec-sing-1}, we require $\calosymbol{d}<800\lambda$, i.e., $\calosymbol{d}<80\ \text{m}$ when $f=3\ \text{GHz}$, and $\calosymbol{d}<8\ \text{m}$ when $f=30\ \text{GHz}$. This requirement is easily met in indoor environments.
}
to assist in range information retrieval \cite{fromenteze2017single,imani2020review} and enable high cross-range resolutions \cite{sheen2001three,jiang2023near}.
The UE moves around the ROI, transmitting pilots and orthogonal frequency division multiplexing (OFDM) communication signals to the AP during different intervals\footnote{
We focus on the uplink paths considering the high computational capabilities of the processing unit, but the discussions can be extended to downlink paths.
},
where the target in the ROI scatters the incident waves.
We assume that the RIS and AP locations, and RIS phase shifts are available at the processing unit.
The UE location can be accurately estimated with advanced RIS \cite{rahal2021ris}, UWB \cite{mahfouz2008investigation}, or RFID \cite{el2018high}-aided localization techniques. We consider only a single UE in this study, but it can be extended to multi-UE scenarios when orthogonal pilot signals are used \cite{huang2023joint}. The generated images of multiple UEs can be fused with crowdsourcing mechanisms \cite{yang2021enabling}.

\subsection{Received Signal Model}

In this study, we realize imaging based on the channel estimation results with known pilots, which do not influence the communication period.
Assume that the UE transmits $K$ pilot symbols for imaging at each position, and $T$ adjacent locations are considered. In each symbol interval, a unique RIS phase shift configuration is used to tune the channel response\footnote{
RIS phase reconfiguration takes about 0.2 us each time \cite{castaldi2021joint}, which is less than the symbol interval (around 1--10 us) in 5G systems \cite{lien20175g}.
}.
When the UE is at the $t$-th position, the $k$-th RIS configuration is used, and the pilot signal $p$ at a single subcarrier is transmitted, the received signal at the AP, which has a dimension of $Q$, is given as \cite{huang2021reconf}
\begin{equation}
\mathbf{r}_{t,k} = (\mathbf{h}^{\rm{LOS}}_{t} + \mathbf{h}^{\rm{ROI}}_{t} + \mathbf{h}^{\rm{RIS}}_{t,k} + \mathbf{h}^{\rm{Ima}}_{t,k} + \mathbf{h}_{t,k}^{\rm{other}}) p + \tilde{\mathbf{z}},
\end{equation}
where $\mathbf{h}^{\rm{LOS}}_{t}$, $\mathbf{h}^{\rm{ROI}}_{t}$, $\mathbf{h}^{\rm{RIS}}_{t,k}$, and $\mathbf{h}^{\rm{Ima}}_{t,k}$ are the channel responses of the UE-AP, UE-ROI-AP, UE-RIS-AP, and UE-ROI-RIS-AP paths, respectively. $\mathbf{h}_{t,k}^{\rm{other}}$ denotes other multipaths that originate from random scatterers and reflectors or experience multiple bounces. $\tilde{\mathbf{z}}$ is additive noise. Among these multipaths,
$\mathbf{h}^{\rm{Ima}}_{t,k}$ experiences the scattering of the ROI and the RIS, involving certain information about the ROI and generating distinct channel responses by varying RIS phase shifts\footnote{
$\mathbf{h}^{\rm{LOS}}_{t}$ and $\mathbf{h}^{\rm{RIS}}_{t,k}$ are not scattered by the ROI, thus involving no direct information about the ROI.
$\mathbf{h}^{\rm{ROI}}_{t}$ is scattered by the ROI, but cannot be reconfigured by the RIS to generate multiple measurements.
Thus, $\mathbf{h}^{\rm{LOS}}_{t}$, $\mathbf{h}^{\rm{RIS}}_{t,k}$, and $\mathbf{h}^{\rm{ROI}}_{t}$ are disturbances to $\mathbf{h}^{\rm{Ima}}_{t,k}$ for the purpose of RIS-aided imaging.
}.
Thus, $\mathbf{h}^{\rm{Ima}}_{t,k}$ can be extracted through channel estimation\footnote{
RIS-reflected paths, including $\mathbf{h}^{\rm{RIS}}_{t,k}$ and $\mathbf{h}^{\rm{Ima}}_{t,k}$, can be extracted by comparing the estimated angle of arrival with the ground truth, calculated using known RIS and AP locations \cite{huang2021reconf}. Assuming the UE antenna is directed towards the ROI during imaging, the energy of $\mathbf{h}^{\rm{RIS}}_{t,k}$ may be minimal, making the extracted RIS-reflected path predominantly $\mathbf{h}^{\rm{Ima}}_{t,k}$.
}
and employed to generate ROI images, as has been used in \cite{jiang2023near,hu2022metasketch,zhu2023ris,sankar2023coded}.
The existence or absence of other multipaths does not have significant impacts on the extraction and utilization of $\mathbf{h}^{\rm{Ima}}_{t,k}$.
The AP with multiple antennas has empowered the extraction of $\mathbf{h}^{\rm{Ima}}_{t,k}$ and can provide multiple measurements for imaging.
However, we only employ the measurement of one antenna for simplicity in the following sections, i.e., only one element in $\mathbf{h}^{\rm{Ima}}_{t,k}$, denoted as $h_{t,k}$, is used for imaging.
The discussions can be easily extended to multi-antenna scenarios by stacking their measurements \cite{zhang20223d}.

\subsection{Measurement Model of the UE-ROI-RIS-AP path}
\label{subsec-channel-model}

As our study does not focus on channel estimation, we assume that $h_{t,k}$ has already been estimated for imaging, as in related literature \cite{jiang2023near,hu2022metasketch,zhu2023ris,sankar2023coded}. Each instance of RIS phase configuration produces a measurement of $h_{t,k}$ with the pilot signal.
We model channel estimation errors as additive Gaussian noise $z_{t,k}$ with the variance of $\chi^2$.
Hence, the measurement of $h_{t,k}$ is given as
\begin{equation}\label{eq-measurement}
y_{t,k} = h_{t,k} + z_{t,k}.
\end{equation}
The objective of imaging is to recover the ROI image from $y_{t,k}$, where $t = 1,2,\ldots, T$, and $k = 1, 2, \ldots, K$.
Next, we establish the relationship between $h_{t,k}$ and the ROI image.

By employing the Born approximation \cite{born2013principles}, we neglect the interactions among different voxels.
Hence, the voxels in Fig. \ref{fig:model-b} act as distinct targets independently.
The subpath originating from the $t$-th UE position, scattered by the $n$-th voxel and the $m$-th RIS element with the $k$-th phase shift configuration, and arriving at the AP can be given as \cite{hu2022metasketch}
\begin{equation}
\label{eq-min}
\begin{aligned}
h_{t,k,n,m} = g\frac{e^{-j2\pi (d_{{\rm{u}}_t,{\rm{v}}_n} + d_{{\rm{s}}_m,{\rm{a}}} + d_{{\rm{s}}_m,{\rm{a}}}) / \lambda}}{(4\pi)^{1.5}d_{{\rm{u}}_t,{\rm{v}}_n} d_{{\rm{s}}_m,{\rm{a}}} d_{{\rm{s}}_m,{\rm{a}}}} e^{-j\omega_{k,m}} x_{t,n},
\end{aligned}
\end{equation}
where $g$ is a constant determined by antenna transmitting and receiving gains.
$\lambda$ is the wavelength, and $\omega_{k,m}$ is the phase shift of the $m$-th RIS element with the $k$-th configuration.
$x_{t,n}\ge0$ is the scattering coefficient of the $n$-th voxel when observed at the $t$-th UE position.
$d_{\star_\circ,\ast_\bullet}$ denotes the distance between the $\circ$-th element of $\star$ and the $\bullet$-th element of $\ast$, where $\star,\ast\in\{\rm{u,v,s,a}\}$ represent the UE, ROI, RIS, and AP, respectively, and $\circ,\bullet\in\{t,n,m\}$.
Summing up the subpaths scattered by all RIS elements and voxels, we have
\begin{equation}\label{eq-large}
\begin{aligned}
{h}_{t,k} & = \sum_{n=1}^N\sum_{m=1}^M h_{t,k,n,m} \\
& = g\mathbf{h}_{{\rm{u}}_t,\rm{v}}^{\rm{T}} \text{diag}(\mathbf{x}_t) \mathbf{H}_{\rm{v,s}} \text{diag}(\boldsymbol{\omega}_k) \mathbf{h}_{\rm{s,a}},
\end{aligned}
\end{equation}
where $\mathbf{x}_t = [x_{t,1}, \ldots, x_{t,N}]^{\rm{T}}$ is the sparse image vector composed of $N$ scattering coefficients with $\operatorname{card}(\operatorname{supp}(\mathbf{x}_t)) \ll N$, as illustrated in Fig. \ref{fig:model-b}.
$\boldsymbol{\omega}_k = [e^{-j\omega_{k,1}}, \ldots, e^{-j\omega_{k,M}}]^{\rm{T}}$ is the vector of RIS phase shift configuration, where $\omega_{k,m} \in [0, 2\pi]$.
\begin{equation}
\mathbf{h}_{\rm{s,a}} = \left[\frac{1}{(4\pi)^{0.5}d_{{\rm{s}}_m,{\rm{a}}}} e^{-j2\pi \frac{d_{{\rm{s}}_m,{\rm{a}}}}{\lambda}}\right]_{M\times 1}
\end{equation}
denotes the free-space subpath from the RIS to the AP.
$\mathbf{h}_{{\rm{u}}_t,{\rm{v}}}\in\mathbb{C}^{N\times 1}$ and $\mathbf{H}_{\rm{v,s}}\in\mathbb{C}^{N\times M}$ are the subpaths from the $t$-th UE position to the ROI and from the ROI to the RIS, respectively, which exhibit the similar form as $\mathbf{h}_{\rm{s,a}}$.
These channels can be calculated with the known locations of the UE, ROI, RIS, and AP.
$\mathbf{h}_{{\rm{u}}_t,{\rm{v}}}$ changes with the UE moving, whereas $\mathbf{H}_{\rm{v,s}}$ and $\mathbf{h}_{\rm{s,a}}$ remain constant since the locations of the ROI, RIS, and AP are fixed.
Combining \eqref{eq-measurement} and \eqref{eq-large}, we have
\begin{equation}\label{htk2}
y_{t,k} = \mathbf{h}_{t,k}^{\rm{T}}\mathbf{x}_t + z_{t,k},
\end{equation}
where
\begin{equation}\label{htk}
\mathbf{h}_{t,k} = g\text{diag}(\mathbf{h}_{{\rm{u}}_t,\rm{v}}) \mathbf{H}_{\rm{v,s}} \text{diag}(\boldsymbol{\omega}_k) \mathbf{h}_{\rm{s,a}}.
\end{equation}
The linear relationship given in \eqref{htk2} is exploited to estimate the ROI image $\mathbf{x}_t$ in Secs. \ref{sec-multi-view} and \ref{sec-case-study}.

\section{Anisotropic Scattering Properties}
\label{sec-anisotropic}

In Sec. \ref{subsec-channel-model}, we have represented voxel scattering coefficients as a function of UE positions.
However, this aspect has not been addressed in prior studies \cite{tong2021joint,tong2022environment}.
In this section, we highlight the importance of incorporating anisotropic scattering characteristics in joint multi-view imaging, considering both RCS properties and diffraction resolution limits.

\begin{figure*}[t]
\begin{equation}
\label{eq-RCS-po}
\sigma_{n'}(A_{\rm{p}}, \lambda, \theta^{\rm{tx}}_{n'}, \theta^{\rm{rx}}_{n'}, \xi_{\rm{px}})=\frac{4 \pi A_{\rm{p}}^{2}}{\lambda^{2}} \cos ^{2}\left(\theta_{n'}^{\rm{tx}}\right) \times \operatorname{Sa}^{2}\left(\frac{\kappa \xi_{\rm{px}}}{2}\left(\sin \left(\theta_{n'}^{\rm{rx}}\right)+\sin \left(\theta_{n'}^{\rm{tx}}\right)\right)\right),
\end{equation}
\hrulefill
\end{figure*}

\begin{figure*}
\centering
\captionsetup{font=footnotesize}
\begin{subfigure}[b]{0.25\linewidth}
\centering
\includegraphics[width=\linewidth]{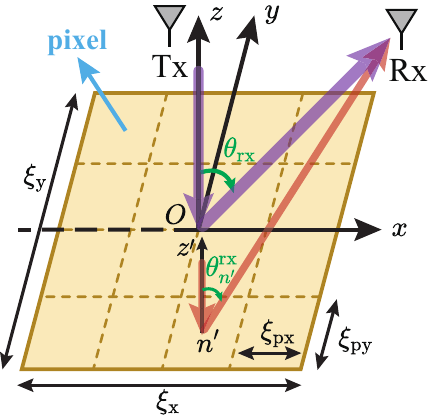}
\caption{}
\label{fig:RCS-a}
\end{subfigure}
\quad\quad
\begin{subfigure}[b]{0.28\linewidth}
\centering
\includegraphics[width=\linewidth]{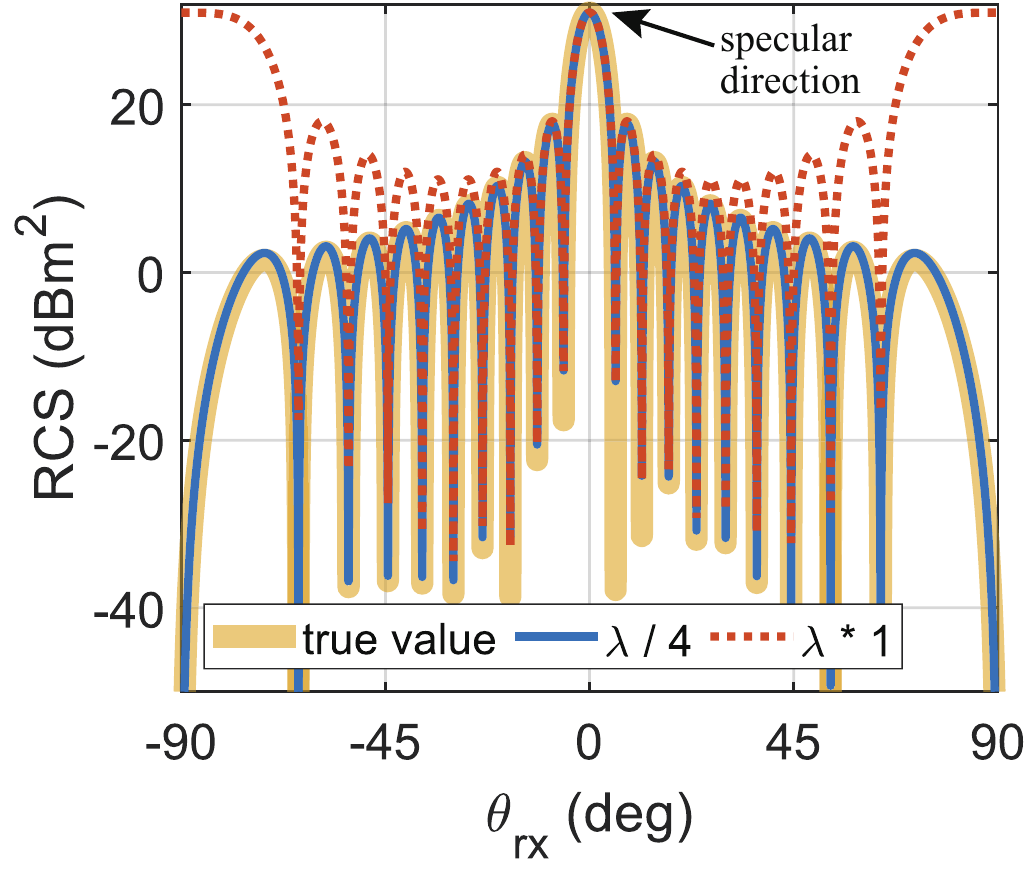}
\caption{}
\label{fig:RCS-b}
\end{subfigure}
\quad\quad
\begin{subfigure}[b]{0.28\linewidth}
\centering
\includegraphics[width=\linewidth]{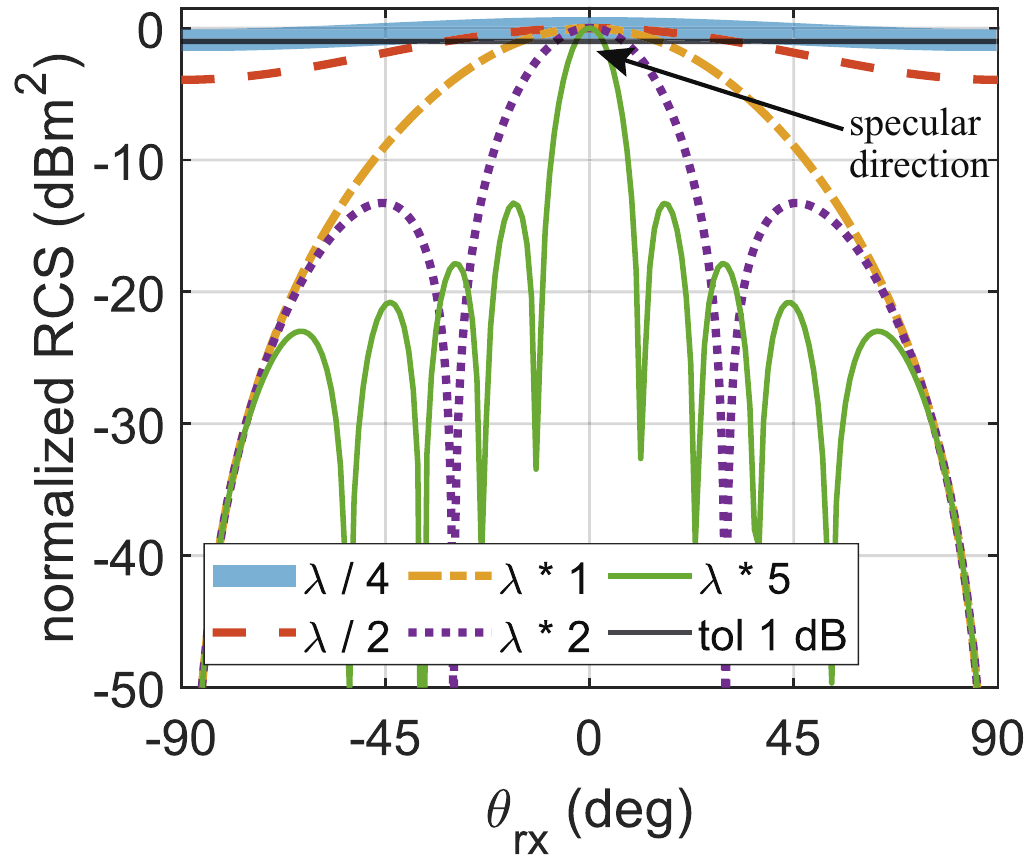}
\caption{}
\label{fig:RCS-c}
\end{subfigure}
\caption{(a) Considered rectangular PEC scatterer; (b) Comparison of RCS values with different pixel sizes ($\lambda / 4$ and $\lambda$) under isotropic assumptions; (c) Normalized RCS values of single pixels versus scattering angles (the legends represent pixel sizes, except that ``tol 1 dB'' denotes the tolerance line of $1\ \rm{dB}$).}
\label{fig:RCS}
\end{figure*}

\subsection{Voxel Size, RCS, and Anisotropic Scattering}
\label{subsec-rcs-halfwavelength}
As depicted in Fig. \ref{fig:model-b}, the ROI has been discretized into $N$ voxels, each of size $\xi_{\rm{vx}} \times \xi_{\rm{vy}} \times \xi_{\rm{vz}}$.
The voxel size plays a crucial role not only in determining the imaging resolution with CS-based imaging methods but also in influencing the scattering properties of the voxels.
As the term ``scattering coefficient'' suggests, voxel scattering properties are closely related to its RCS.
Thus, in this subsection, we utilize the RCS of the ROI, which represents the influence of the ROI on EM propagation environments and can be obtained from near-field ROI images \cite{huang2023joint}, as a metric to evaluate whether the voxels can be considered isotropic or anisotropic.

Considering that 3D target surfaces can be approximated as a collection of 2D planes, we simplify the target as a 2D rectangular perfectly electrical conductor (PEC) scatterer with negligible thickness, positioned in the $xOy$ plane, as illustrated in Fig. \ref{fig:RCS-a}.
The incident wave propagates along the negative direction of the $z$-axis, while the observing direction rotates in the $xOz$ plane.
Analogous to Fig. \ref{fig:model-b}, the rectangular scatterer is divided into $N'$ uniform pixels, each with a size of $A_{\rm{p}} = \xi_{\rm{px}} \times \xi_{\rm{py}}$.
Assuming that the transmitter and receiver are located in the far field of the pixels, the RCS of the $n'$-th pixel with the method of physical optics is expressed as \eqref{eq-RCS-po} at the top of the next page \cite{huang2023joint,Merrill1990radar,balanis2012advanced}, where $\text{Sa}(x) = \sin x/x$ with $\text{Sa}(0)=1$, and $\kappa=2\pi/\lambda$ is the wavenumber.
$\xi_{\rm{px}}$ is the pixel length along the $x$-axis.
$\theta_{n'}^{\rm{tx}}$ and $\theta_{n'}^{\rm{rx}}$ denote the elevation angles pointing from the pixel center to the transmitter and receiver, respectively, as depicted in Fig. \ref{fig:RCS-a}.
The pixel RCS given in \eqref{eq-RCS-po} exhibits anisotropic properties with respect to the observation angles $\theta^{\rm{tx}}_{n'}$ and $\theta^{\rm{rx}}_{n'}$.
Although the RCS is traditionally defined in the far field, its applicability has been extended to the near-field region in literature \cite{deban2009deterministic}, as a function of both distances and observing angles.
The general RCS is defined by summing up the RCS values of all pixels while considering the phase shifts resulting from interpixel path length differences, given as \cite{huang2023joint,Merrill1990radar,balanis2012advanced}
\begin{equation}\label{eq-general-rcs}
\sigma = \left| \sum_{n' = 1}^{N'} \sqrt{\sigma_{n'}} e^{\frac{-j2\pi \left(\calosymbol{d}_{n'}^{\rm{tx}} + \calosymbol{d}_{n'}^{\rm{rx}}\right)}{\lambda}} \right|^2,
\end{equation}
where $\calosymbol{d}_{n'}^{\rm{tx}}$ and $\calosymbol{d}_{n'}^{\rm{rx}}$ are the distances from the $n'$-th pixel to the transmitter and receiver, respectively.

Next, we study the scattering properties of pixels through simulations.
The working frequency is set to $f = 3 \ \rm{GHz}$, rectangular plate size is $\xi_{\rm{x}} = \xi_{\rm{y}} = 10 \lambda$, and pixel size satisfies $\xi_{\rm{px}} = \xi_{\rm{py}}$.
In Fig. \ref{fig:RCS-b}, we compare the calculated scatterer RCS values using \eqref{eq-general-rcs} with different pixel sizes.
The RCS values of pixels are obtained from \eqref{eq-RCS-po} with $\sigma_{n'}(A_{\rm{p}}, \lambda, \theta^{\rm{tx}}_{n'} = 0^\circ, \theta^{\rm{rx}}_{n'} = 0^\circ, \xi_{\rm{px}})$, under the assumption of isotropic scattering.
Instead, the true values are derived by taking into account the anisotropic RCS of each pixel as a function of $\theta^{\rm{tx}}_{n'}$ and $\theta^{\rm{rx}}_{n'}$.
The RCSs derived from isotropic pixels with $\xi_{\rm{px}} = \lambda / 4$ closely match the true values.
However, significant discrepancies arise when treating the pixels with the size of $\lambda\times \lambda$ as isotropic.
Further observations are illustrated in Fig. \ref{fig:RCS-c}, where different pixel sizes exhibit distinct normalized RCS fluctuations with respect to the observation angles.
The overall fluctuations of the pixel with $\xi_{\rm{px}} = \lambda / 4$ within the range $[-90^\circ, 90^\circ]$ are less than $1\ \rm{dB}$, making it reasonably considered isotropic.
As the pixel size increases, RCS fluctuations also increase.
When the pixel sizes exceed $\lambda/2$, significant fluctuations occur, indicating the dominance of anisotropic properties.
Extrapolating these results to 3D voxels, we can conclude that anisotropic scattering becomes significant when voxel sizes are larger than $\lambda / 2$, whereas voxels can be considered isotropic when their sizes are much smaller than $\lambda / 2$.
The next subsection examines whether the voxel sizes can be selected to be less than $\lambda / 2$, allowing for the applicability of the isotropic assumption.

\subsection{Diffraction Resolution Limits}
\label{sec-diffraction}

Diffraction effects impose limitations on the maximum achievable resolution in radio imaging.
Assuming that the target is located in front of the center of the RIS, the diffraction resolution limits are given as \cite{sheen2001three,jiang2023near}
\begin{equation}\label{eq-diffraction}
\delta_{\rm{range}} = \frac{c}{2B}, \quad \delta_{\rm{cross-range}} = \frac{\lambda}{2\sin(\psi / 2)},
\end{equation}
where $\delta_{\rm{range}}$ and $\delta_{\rm{cross-range}}$ are the resolution limits in range and cross-range directions (along the $x$-axis and $y$- or $z$-axis as shown in Fig. \ref{fig:model-a}), respectively.
$c$ is the speed of light, and $B$ is the bandwidth.
$\psi$ is the angle subtended by the RIS aperture.
Note that the range resolution limits in \eqref{eq-diffraction} are widely used in near-field SAR imaging for performance evaluation, despite being originally derived in the far field of imaging apertures \cite{imani2020review}.
Additionally, let $D$ denote the distance between the ROI and the RIS aperture, and $R$ represent the size of the RIS aperture. Using these parameters, we can derive
\begin{equation}\label{eq-subtended-angle}
\sin(\psi / 2) = \sqrt{\frac{(R / 2)^2}{(R / 2)^2 + D^2}} = \frac{R}{\sqrt{R^2 + 4D^2}}.
\end{equation}
Thus, closer distances and larger RIS array sizes can improve the cross-range resolution.
The diffraction resolution limit provides an assessment of the imaging capabilities. Systems with smaller resolution limit values are expected to achieve higher imaging performance.

In this subsection, we explore the relationship between diffraction resolution limits and the $\lambda / 2$ boundary derived in Sec. \ref{subsec-rcs-halfwavelength}.
First, we consider the range resolution limit, which requires a bandwidth of $B = f$ when $\delta_{\rm{range}} = \lambda / 2$.
Here, $f = \lambda / c$ represents the subcarrier frequency.
However, achieving the resolution of $\lambda / 2$ in the range direction is impractical because communication systems typically cannot provide such a large bandwidth.
Second, we have $\delta_{\rm{cross-range}} > \lambda / 2$ due to the fact that $\sin(\psi / 2) < 1$, as indicated by \eqref{eq-subtended-angle}.
Therefore, the range and cross-range resolutions in the considered imaging scenarios have been limited to be larger than $\lambda / 2$.

\begin{remark}
In conclusion, voxels can only be considered isotropic when their sizes are significantly smaller than $\lambda / 2$, whereas the achievable resolutions in imaging systems are typically larger than $\lambda / 2$.
Since voxel sizes correspond to the desired resolutions in CS-based imaging algorithms, they should be larger than $\lambda / 2$.
Therefore, it is crucial to consider anisotropic scattering characteristics in joint multi-view imaging, especially when voxel sizes are electronically large.
This presents a challenge in joint multi-view imaging as the observed images at different UE positions vary.
However, sequential images obtained along a continuous trajectory maintain high levels of temporal correlations. The following section focuses on joint multi-view imaging by exploiting the underlying correlations.
\end{remark}

\section{Joint Multi-view Single-frequency 3D Imaging}
\label{sec-multi-view}

High-resolution images require large RIS aperture sizes and close distances from the ROI to the RIS, as indicated in Sec. \ref{sec-diffraction}. However, this section aims to mitigate these requirements by introducing a novel joint multi-view imaging scheme that leverages the temporal correlations in multi-view images. Specifically, we assume the UE moves step by step around the ROI, and the RIS phase shift configurations are randomly altered at each step to capture the scattering features of the target. The EM-turbo-GAMP algorithm captures the mathematically modeled variations in sequential images which are caused by occlusion effects and anisotropic scattering.

\subsection{Mathematic Multi-view Image Model}
\label{sec-sub-mmv-model}

According to \eqref{htk2}, we stack the measurements with $K$ distinct RIS phase shifts and derive
\begin{equation}\label{eq-y=Ax}
\mathbf{y}_{t} = \mathbf{A}_{t}\mathbf{x}_{t} + \mathbf{z}_{t}, \quad t = 1, 2, \ldots, T,
\end{equation}
where $\mathbf{y}_{t} = [y_{t,1}, y_{t,2}, \ldots, y_{t,K}]^{\rm{T}} \in \mathbb{C}^{K\times 1}$ is the extracted UE-ROI-RIS-AP paths that originate from the $t$-th UE position.
$\mathbf{x}_{t} \in \mathbb{R}^{N\times 1}$ donates the observed image at the $t$-th UE position.
$\mathbf{z}_{t} = [z_{t,1}, z_{t,2}, \ldots, z_{t,K}]^{\rm{T}} \in \mathbb{C}^{K\times 1} $ is additive zero-mean Gaussian white noise with covariance matrix $\chi^2\mathbf{I}$, where $\mathbf{I}$ is the identity matrix.
The sensing matrix is given as
\begin{equation}\label{eq-new-A}
\mathbf{A}_{t} = [\mathbf{h}_{t,1}, \mathbf{h}_{t,2}, \ldots, \mathbf{h}_{t,K}]^{\rm{T}} = \boldsymbol{\Omega}\mathbf{B}_t,
\end{equation}
where $\boldsymbol{\Omega} = \left[\boldsymbol{\omega}_1, \boldsymbol{\omega}_2, \ldots, \boldsymbol{\omega}_K\right]^{\rm{T}} \in \mathbb{C}^{K\times M}$, and
\begin{equation}
\mathbf{B}_{t} = g{\rm{diag}}(\mathbf{h}_{\rm{s,a}})\mathbf{H}_{\rm{v,s}}^{\rm{T}}{\rm{diag}}(\mathbf{h}_{{\rm{u}}_t, {\rm{v}}}).
\end{equation}
Note that \eqref{eq-y=Ax} represents a combination of $T$ single-view imaging problems at adjacent UE positions, fundamentally differing from the sensing matrix perturbation model used in \cite{tong2022environment}.
$\{\mathbf{x}_{t}\}_{t=1}^T$ includes simultaneous variations in supports and amplitudes, and $\mathbf{A}_t$ varies from one position to another.
Moreover, \eqref{eq-y=Ax} exhibits an underdetermined nature at each UE position, since a large ROI can be divided into hundreds or thousands of voxels, while the number of RIS configurations $K \ll N$.
Therefore, solving the problem described by \eqref{eq-y=Ax} is challenging.
Previous studies have successfully reconstructed 2D images by exploiting the sparse properties of the ROI at a single UE position \cite{hu2022metasketch,zhu2023ris}.
However, in this section, our goal is to achieve joint multi-view single-frequency 3D imaging by simultaneously harnessing the sparse properties and temporal correlations presented in $\{\mathbf{x}_{t}\}_{t=1}^T$.

First, we model the distribution of $\mathbf{x}_{t}$ considering its sparsity, which is evident as depicted in Fig. \ref{fig:model-b}, resulting that the majority of the elements in $\mathbf{x}_{t}$ are zero.
In this study, we assume that the distribution of voxel scattering coefficients follows a Bernoulli-Gaussian distribution, whose probability density function (PDF) is given as \cite{tong2021joint}
\begin{equation}\label{eq-bg-pdf}
p_{\mathcal{X}_{t,n}}\left({x}_{t,n} ; \alpha, \eta, \varsigma^{2}\right)=(1-\alpha) \delta\left({x}_{t,n}\right)+\alpha \mathcal{N}\left({x}_{t,n}; \eta, \varsigma^{2}\right),
\end{equation}
where $\alpha = \operatorname{card}\left(\operatorname{supp}(\mathbf{x}_t)\right)/N$ denotes the sparse rate of $\mathbf{x}_{t}$, and $\delta(\cdot)$ is the Dirac delta function.
$\mathcal{N}\left(\cdot; \eta, \varsigma^{2}\right)$ represents the PDF of a Gaussian distribution with mean $\eta$ and variance $\varsigma^2$.

Next, our attention shifts to the temporal correlations presented in $\{\mathbf{x}_{t}\}_{t=1}^T$, which are observed at adjacent UE positions along a continuous trajectory.
To describe these correlations, we introduce two stochastic processes, $\left\{\mathbf{s}_{t}\right\}_{t=1}^{T}$ and $\left\{\mathbf{a}_{t}\right\}_{t=1}^{T}$, as proposed in \cite{ziniel2013dynamic}.
Here, $\mathbf{s}_{t}\in\{0,1\}^N$ characterizes the slowly varying support set of $\mathbf{x}_{t}$, and $\mathbf{a}_{t}\in\mathbb{R}^N$ accounts for the smooth evolution of non-zero coefficients in $\mathbf{x}_{t}$.
With these processes, we can express the relationship as
\begin{equation}
{x}_{t,n} = {s}_{t,n} {a}_{t,n},
\end{equation}
where ${s}_{t,n}$ and ${a}_{t,n}$ are the $n$-th element of $\mathbf{s}_{t}$ and $\mathbf{a}_{t}$, respectively.
Consequently, ${x}_{t,n}=0$ when ${s}_{t,n}=0$, and ${x}_{t,n}={a}_{t,n}$ when ${s}_{t,n}=1$.

\begin{figure*}
    \centering
    \includegraphics[width=0.65\textwidth]{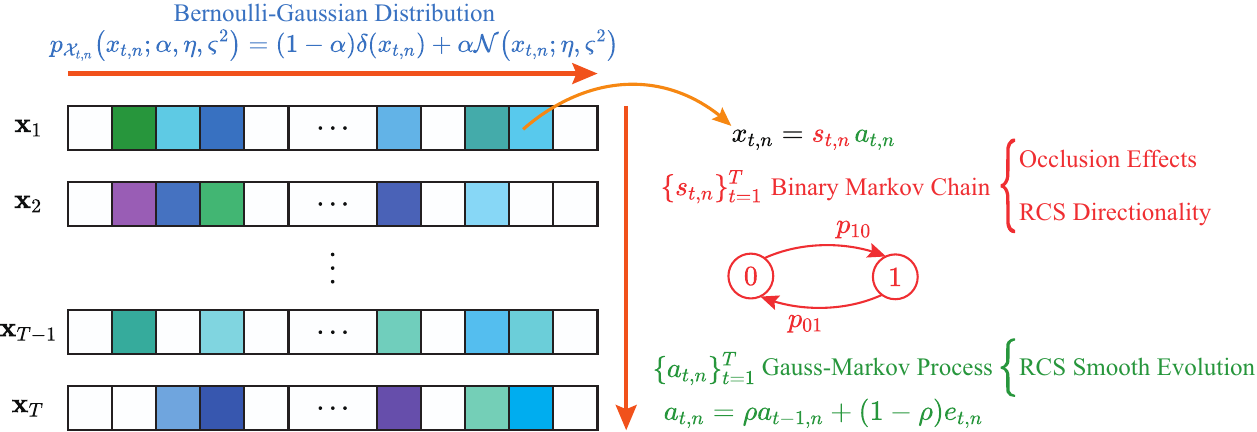}
    \captionsetup{font=footnotesize}
    \caption{Illustration of the proposed multi-view image model considering occlusion effects and anisotropic scattering, where white squares represent zero coefficients, and colorful squares denote varying non-zero coefficients.}
    \label{fig:mmv-model}
\end{figure*}

\textbf{Slow variation of image supports $\left\{\mathbf{s}_{t}\right\}_{t=1}^{T}$.}
As the UE slowly moves around the ROI, the visible voxels of the UE exhibit gradual variations.
This variation originates from occlusion effects and RCS properties.
On the one hand, the voxels in the ROI may occlude each other \cite{tong2022environment}, causing some voxels invisible.
At the current UE position, the corresponding scattering coefficients of these voxels are set to zero.
On the other hand, voxel scattering exhibits a certain directionality.
For example, significant scattered energy can be observed in the specular directions of incident waves (when scattering elevation angles are equal to the incident ones).
However, as the scattering angles deviate from this direction, the received signal energy decays, making non-zero voxels less visible in certain directions due to their low scattered energy that may be submerged in background noise, as illustrated in Figs. \ref{fig:RCS-b} and \ref{fig:RCS-c}.
These voxels are considered to possess zero scattering coefficients at the present UE position.
The variation of the support for the $n$-th voxel $\{{s}_{t,n}\}_{t=1}^T$ with the movement of the UE is modeled as a binary Markov chain, where the transition probabilities are given by
\begin{subequations}
\begin{align}
p_{01} \triangleq \mathrm{P}\left\{{s}_{t,n}=0 \mid {s}_{t-1,n}=1\right\},\\
p_{10} \triangleq \mathrm{P}\left\{{s}_{t,n}=1 \mid {s}_{t-1,n}=0\right\}.
\end{align}
\end{subequations}
Meanwhile, we assume that the Markov chain for each voxel operates in steady states, thereby obtaining ${\rm{P}}\left\{{\rm{s}}_{t,n}=1\right\}=\alpha$.
Consequently, we can express
\begin{equation}
p_{10}=\frac{\alpha p_{01}}{1-\alpha}.
\end{equation}
Due to the sparsity of $\mathbf{x}_{t}$ and the slow variation of signal supports $\{{s}_{t,n}\}_{t=1}^T$, $\{\mathbf{x}_{t}\}_{t=1}^T$ can be considered roughly joint sparse, satisfying
\begin{equation}
\operatorname{card}\left(\bigcup_{t=1}^{T} \operatorname{supp}\left(\mathbf{x}_{t}\right)\right)=S \ll N,
\end{equation}
where $\bigcup$ computes the union of sets, and $S$ represents the number of observed non-zero voxels.

\textbf{Smooth evolution of coefficients $\left\{\mathbf{a}_{t}\right\}_{t=1}^{T}$.}
As illustrated in Fig. \ref{fig:RCS-c}, the RCS values corresponding to different voxel sizes exhibit smooth variations with respect to the observation angles.
Therefore, the coefficients of visible non-zero voxels undergo a smooth evolution as the UE moves along a continuous trajectory.
In this study, we model the coefficient evolution $\{{a}_{t,n}\}_{t=1}^T$ of the $n$-th voxel with slowly moving UE positions as a Gauss-Markov stochastic process, which can be represented as
\begin{equation}
{a}_{t,n}=\rho {a}_{t-1,n}+(1-\rho) {e}_{t,n},
\end{equation}
where $\rho$ is the temporal correlation parameter, and ${e}_{t,n}$ is a perturbation random Gaussian variable with the mean and variance being $\eta_{\rm{e}}$ and $\varsigma^2_{\rm{e}}$, respectively.
According to the mean and variance constraints of Bernoulli-Gaussian distribution given in \eqref{eq-bg-pdf}, we have
\begin{equation}
\eta_{e}=\eta, \quad \varsigma_{e}^{2}=\frac{1+\rho}{1-\rho} \varsigma^{2}.
\end{equation}

The support variation and coefficient evolution models proposed above have been summarized in Fig. \ref{fig:mmv-model}.
Despite the variations in image supports and coefficients, $\{\mathbf{x}_{t}\}_{t=1}^T$ exhibits high degrees of temporal correlations that can be exploited to enhance imaging performances.

\begin{remark}
A voxel can only be observed when it lies within the LOS regions of both the UE and the RIS simultaneously.
In the case of a stationary ROI, the visible voxels of the fixed RIS remain constant.
However, the UE may only perceive a subset of these voxels, and the observed voxels at different positions overlap and complement each other.
Thus, joint multi-view imaging allows for the maximization of the sensed voxels.
Furthermore, the use of multiple RISs can further expand the imaged regions.
However, for the sake of simplicity, this study focuses on the single-RIS scenario depicted in Fig. \ref{fig:model-a}.
\end{remark}

\subsection{Joint Multi-view Imaging Algorithm}

In this subsection, our objective is to solve the problem presented in \eqref{eq-y=Ax} and achieve joint multi-view imaging by leveraging the temporal correlations of $\{\mathbf{x}_t\}_{t=1}^T$.
Rather than fusing the imaging results of the $T$ single-view images, we propose to jointly estimate the $T$ images.
Based on \cite{bellili2019generalized}, the minimum mean square error (MMSE) estimates of $x_{t,n}$ can be obtained as follows:
\begin{equation}
\begin{aligned}
\hat{{x}}_{t,n} & =\underset{\widehat{{x}}_{t,n}}{\operatorname{argmin}}\  \mathbb{E}_{{\mathcal { X }}_{t,n}, \overline{\boldsymbol{\mathcal{Y}}}}\left\{\left({x}_{t,n}-\hat{{x}}_{t,n}\right)^{2}\right\} \\
& =\mathbb{E}_{{\mathcal { X }}_{t,n} \mid \overline{\boldsymbol{\mathcal{Y}}}}\left\{{x}_{t,n} \mid \overline{\mathbf{y}}\right\},
\end{aligned}
\end{equation}
where $\overline{\mathbf{y}} \triangleq \left[\mathbf{y}_{1}; \mathbf{y}_{2}; \ldots; \mathbf{y}_{T}\right]$.
Thus, the problem given in \eqref{eq-y=Ax} can be reformulated as the computation of the marginal PDF $p_{{\mathcal { X }}_{t,n} \mid \overline{\boldsymbol{\mathcal{Y}}}}\left({x}_{t,n} \mid \overline{\mathbf{y}}\right)$.
However, this calculation is analytically intractable and computationally prohibitive.
To address this challenge, we employ the Bayesian inference framework, approximating the marginal PDF by the product of tractable PDFs associated with small subsets of hidden variables and parameters.
\begin{figure*}[t]
\begin{equation}
\label{aaa}
p(\overline{\mathbf{x}}, \overline{\mathbf{s}}, \overline{\mathbf{a}} \mid \overline{\mathbf{y}} ; \mathcal{P})  \propto \prod_{t=1}^{T}\left(\prod_{k=1}^{K} p\left({y}_{t,k} \mid \mathbf{x}_{t} ; \mathcal{P}\right) \prod_{n=1}^{N} p\left({x}_{t,n} \mid {s}_{t,n}, {a}_{t,n} ; \mathcal{P}\right) p\left({s}_{t,n} \mid {s}_{t-1,n} ; \mathcal{P}\right) p\left({a}_{t,n} \mid {a}_{t-1,n} ; \mathcal{P}\right)\right),
\end{equation}
\hrulefill
\end{figure*}
Given the measurements $\overline{\mathbf{y}}$, the joint posterior distribution of all random variables (including the signals, supports, and amplitudes) can be decomposed as \eqref{aaa} at the top of the next page, where the symbols of random variables are omitted for simplicity.
$\mathcal{P}=\left\{\chi^{2}, \alpha, \eta, \varsigma^{2}, p_{01}, \rho\right\}$ denotes the modeling parameters presented in Sec. \ref{sec-sub-mmv-model}.
However, simultaneously calculating the unknown PDFs in \eqref{aaa} and leveraging the underlying correlations in the sequential images are extremely complex tasks that traditional algorithms like the GAMP algorithm \cite{rangan2011generalized} may struggle to handle.

\begin{figure*}[t]
    \centering
    \includegraphics[width=0.7\textwidth]{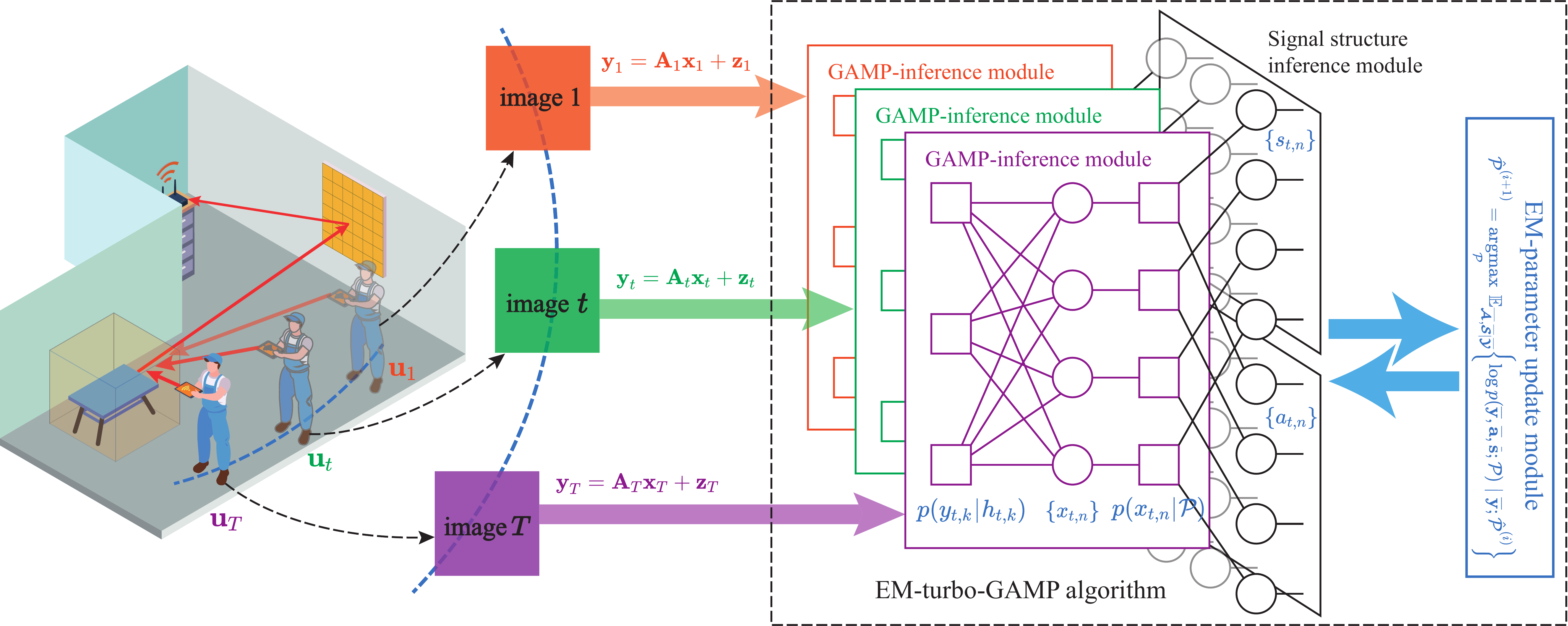}
    \captionsetup{font=footnotesize}
    \caption{Illustration of the EM-turbo-GAMP algorithm for joint multi-view imaging, where the UE moves along a continuous trajectory (circles and squares in the right part of this figure represent random variables and probability distribution factors, respectively).}
    \label{fig:mmv-algorithm}
\end{figure*}

\begin{algorithm}[t]
    \caption{EM-turbo-GAMP algorithm \cite{ziniel2012generalized}.}
    \label{ag-em}
    \begin{algorithmic}[1]
            \State $\mathbf{input:}$ $\{\mathbf{A}_t\}_{t=1}^T$, $\{\mathbf{y}_t\}_{t=1}^T$, priors $\mathcal{P}^{(0)}$ and $\boldsymbol{\aleph}^{(0)}$, and maximum iteration number $I_{\rm{max}}$.

            \State $\mathbf{for} \  i = 1 \ {\rm{to}} \  I_{\rm{max}} \ \mathbf{do}$

            \State \hspace{0.5cm} $\mathbf{for} \  t = 1 \ {\rm{to}} \  T \ \mathbf{do}$

            \State \hspace{1.0cm} Execute the GAMP \cite{rangan2011generalized} algorithm with $\mathbf{A}_t$, $\mathbf{y}_t$,

            \hspace{0.8cm} $\mathcal{P}^{(i-1)}$, and $\boldsymbol{\aleph}^{(i-1)}$, deriving $\mathbf{x}_t^{(i)}$.

            \State \hspace{0.5cm} $\mathbf{end\ for}$

            \State \hspace{0.5cm} Explore probabilistic structures lying in $\{\mathbf{x}^{(i)}_t\}_{t=1}^T$

            \hspace{0.4cm} and derive $\boldsymbol{\aleph}^{(i)}$.

            \State \hspace{0.5cm} Derive $\mathcal{P}^{(i)}$ according to \eqref{eq-emlearning}.

            \State $\mathbf{end\ for}$

            \State $\mathbf{output:}$ the estimated images $\{{\mathbf{x}}_t^{(I_{\rm{max}})}\}_{t=1}^T$.

    \end{algorithmic}
\end{algorithm}

In this study, we utilize the EM-turbo-GAMP algorithm proposed in \cite{ziniel2012generalized} to jointly estimate the multi-view images $\left\{\mathbf{x}_{t}\right\}_{t=1}^{T}$.
The algorithm consists of three modules: the GAMP-inference module, the signal structure inference module, and the EM-parameter update module, which are illustrated in Fig. \ref{fig:mmv-algorithm} using factor graphs.
In this algorithm, $\mathcal{P}$ and $\boldsymbol{\aleph} = \{\left\{\mathbf{s}_{t}\right\}_{t=1}^{T},\left\{\mathbf{a}_{t}\right\}_{t=1}^{T}\}$ are first initialized with available prior knowledge.
Then, the GAMP-inference module employs the GAMP algorithm \cite{rangan2011generalized} to respectively derive the estimates of the marginal PDFs $\{p_{{\mathcal{X}}_{t,n} \mid \boldsymbol{\mathcal{Y}}_{t}}\left({x}_{t,n} \mid \mathbf{y}_{t}\right)\}_{t=1}^T$ and unknown variables $\{\mathbf{x}_{t}\}_{t=1}^T$ for the $T$ linear inverse problems.
Then, the image correlations lying in $\{\mathbf{x}_{t}\}_{t=1}^T$ are explored to update the hidden variables $\boldsymbol{\aleph}$, which are fed back to help the estimation of the GAMP algorithm.
The images $\{\mathbf{x}_{t}\}_{t=1}^T$ are progressively refined through iterations.
During each iteration, the unknown parameters $\mathcal{P}$ are updated by solving the following problem with the EM learning algorithm \cite{dempster1977maximum}
\begin{equation}\label{eq-emlearning}
\hat{\mathcal{P}}^{(i + 1)}=\underset{\mathcal{P}}{\operatorname{argmax}} \ \mathbb{E}_{\overline{\boldsymbol{\mathcal{A}}}, \overline{\boldsymbol{\mathcal{S}}} \mid \overline{\boldsymbol{\mathcal{Y}}}} \left\{\log p(\overline{\mathbf{y}}, \overline{\mathbf{a}}, \overline{\mathbf{s}} ; \mathcal{P}) \mid \overline{\mathbf{y}} ; \hat{\mathcal{P}}^{(i)}\right\},
\end{equation}
where $\hat{\mathcal{P}}^{(i)}$ is the estimate of $\mathcal{P}$ in the $i$-th iteration.
According to \cite{ziniel2013dynamic}, closed-form solutions of \eqref{eq-emlearning} can be derived by decoupling the PDF $p(\overline{\mathbf{y}}, \overline{\mathbf{a}}, \overline{\mathbf{s}} ; \mathcal{P})$ into tractable PDFs, which are obtained in the main inference procedure.
The EM-turbo-GAMP algorithm is summarized in Algorithm \ref{ag-em}.
According to \cite{ziniel2013dynamic}, the primary computational complexity of the EM-turbo-GAMP algorithm originates from the execution of the GAMP algorithm and is linear with respect to $T$, $K$, and $N$.

\begin{remark}
The proposed joint multi-view imaging scheme provides benefits by leveraging the temporal correlations lying in $\{\mathbf{x}_{t}\}_{t=1}^T$.
The correlations are characterized by $p_{01}$ and $\rho$ and influenced by the UE step length $d_0$ between adjacent UE positions.
When $d_0$ is extremely small, the adjacent positions can be approximately categorized as a single point, and the observed images are nearly identical.
Consequently, multi-view imaging is reduced to single-view imaging.
As $d_0$ gradually increases, the perceived images at adjacent UE positions may exhibit significant differences while still maintaining temporal correlations.
The correlations can be captured by exploring signal structures to enhance imaging performance, as proposed in this study.
However, when $d_0$ becomes very large, the temporal correlation in $\{\mathbf{x}_{t}\}_{t=1}^T$ may become too weak to provide meaningful information, and joint multi-view imaging is essentially equivalent to $T$ separate single-view imaging tasks.
Therefore, the proposed joint multi-view imaging scheme requests a proper step length $d_0$ to fully exploit image correlations.
\end{remark}

\begin{remark}
The proposed joint multi-view imaging scheme offers several advantages over the previous studies \cite{sheen2001three,tong2022environment,wu2021through,hu2022metasketch,jiang2023near,zhu2023ris,sankar2023coded}.
Compared to traditional SAR imaging techniques \cite{sheen2001three}, the proposed method leverages rapid and maneuverable RIS phase shift tuning \cite{castaldi2021joint} to obtain multiple measurements, enabling real-time imaging.
Unlike the sensing matrix perturbation model in \cite{tong2022environment} and the joint sparse model in \cite{wu2021through}, the proposed scheme considers both occlusion effects and anisotropic scattering, providing a more realistic depiction of scattering properties.
As an extension of prior studies \cite{hu2022metasketch,jiang2023near,zhu2023ris,sankar2023coded}, this study emphasizes the temporal correlations in images observed at adjacent UE positions.
Multi-view observations help retrieve range and cross-range information, resulting in significantly enhanced 3D imaging performances with low measurement overhead.
Furthermore, the proposed joint multi-view imaging method obtains an angular scattering response for each voxel, allowing for the characterization of anisotropy.
Such information can be a crucial attribute for target recognition and scene understanding \cite{ccetin2014sparsity}.
\end{remark}

\section{Theoretical Analysis of Imaging Performance}
\label{sec-case-study}

In Section \ref{sec-multi-view}, the 3D imaging performances have been improved by exploiting the temporal correlations presented in sequential images.
However, the performance of joint multi-view imaging is inherently linked to that of the single-view imaging at each UE position.
In this section, we simplify the problem and focus on single-view imaging to investigate the geometric properties that limit imaging performances.
We also explore the benefits of optimizing RIS phase shifts.

\begin{table}[t]
\renewcommand{\arraystretch}{1.5}
  \centering
  \fontsize{8}{8}\selectfont
  \captionsetup{font=small}
  \captionof{table}{System parameters of the studied cases.}\label{tab-case-study}
  \begin{threeparttable}
    \begin{tabular}{c|c}
      \specialrule{1pt}{0pt}{-1pt}\xrowht{12pt}
      Parameters & Values \\
      \hline
      ROI size & $20\lambda \times 20\lambda \times 20\lambda$ \\
      Voxel number & $N = 10\times10\times10$ \\
      Voxel size & $\xi_{\rm{v}} = 2\lambda$ \\
      Imaging distance & $D = 50\lambda$ \\
      RIS element number & $M = 48 \times 48$ \\
      RIS element size & $\xi_{\rm{s}} = \lambda / 2$ \\
      \specialrule{1pt}{0pt}{0pt}
    \end{tabular}
  \end{threeparttable}
\end{table}

\begin{figure*}
    \centering
    \captionsetup{font=footnotesize}
    \begin{subfigure}[b]{0.28\linewidth}
    \centering
    \includegraphics[width=\linewidth]{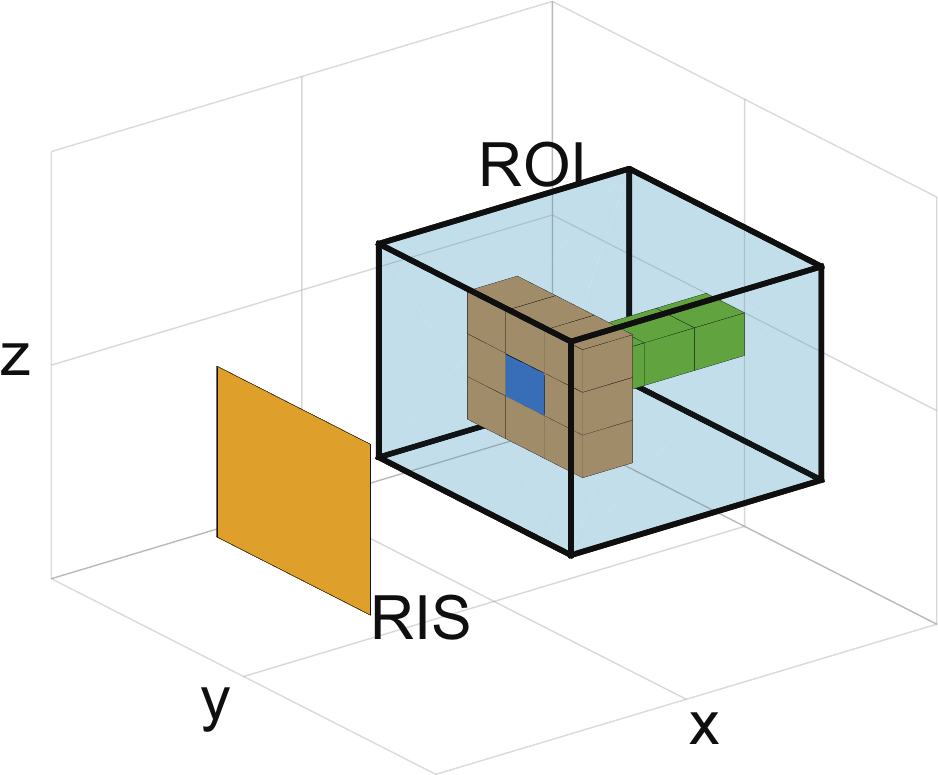}
    \caption{}
    \label{fig:subpath-a}
    \end{subfigure}
    \quad
    \begin{subfigure}[b]{0.28\linewidth}
    \centering
    \includegraphics[width=\linewidth]{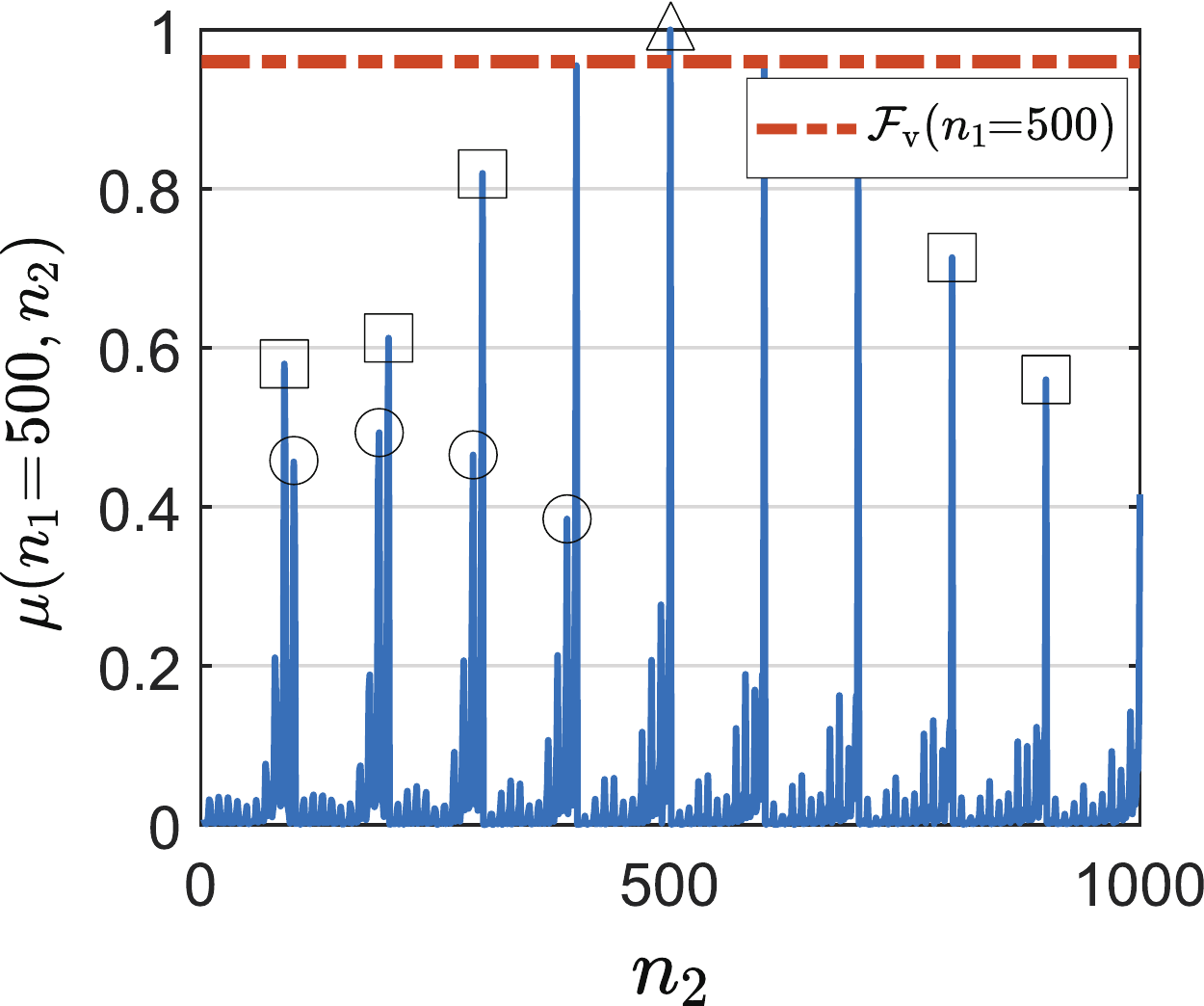}
    \caption{}
    \label{fig:subpath-b}
    \end{subfigure}
    \quad
    \begin{subfigure}[b]{0.28\linewidth}
    \centering
    \includegraphics[width=\linewidth]{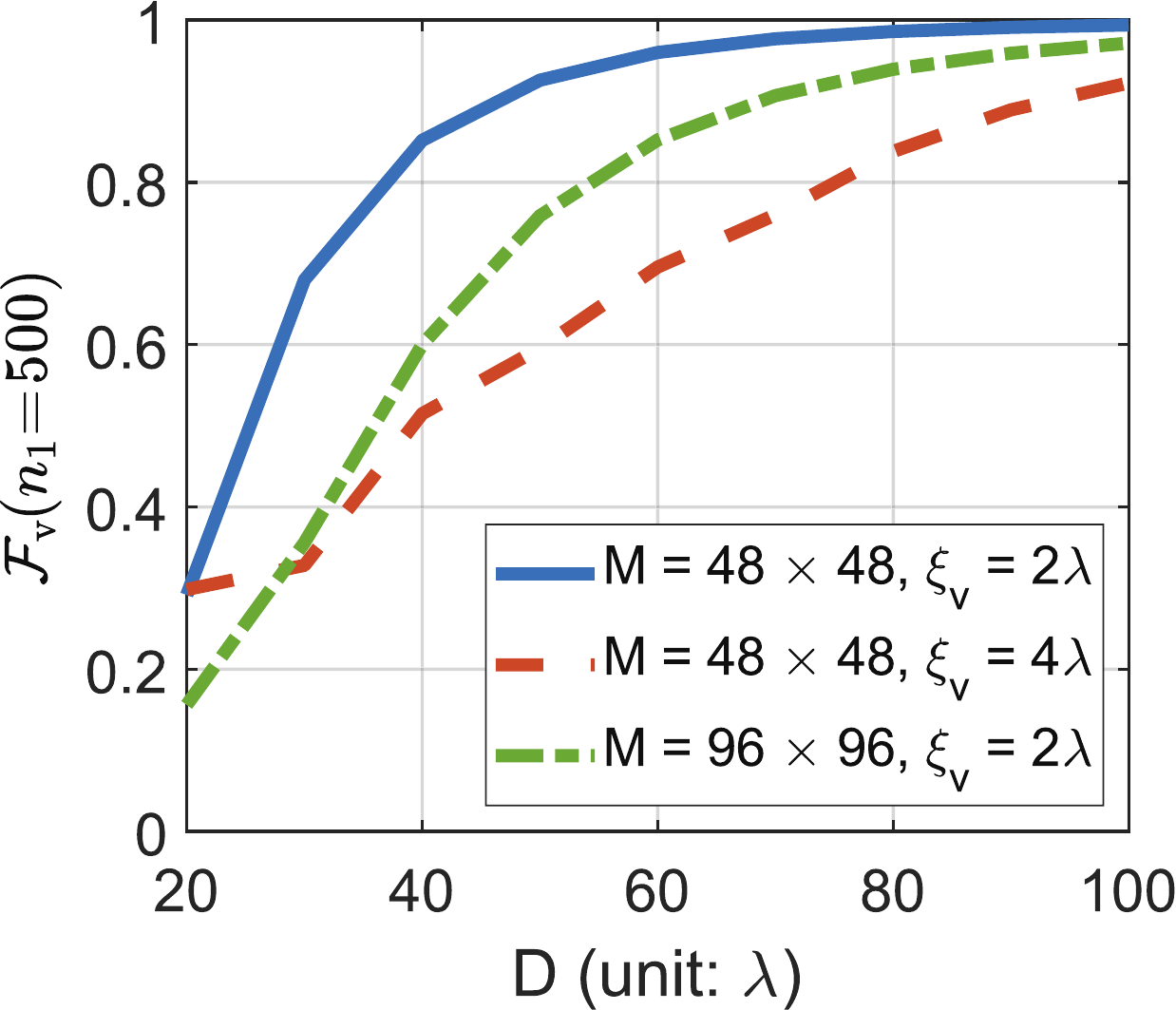}
    \caption{}
    \label{fig:subpath-c}
    \end{subfigure}
    \caption{(a) The imaging scenario, where the blue, green, and brown voxels denote the considered voxel, and those in the range and cross-range directions, respectively; (b) Subpath correlation with $n_1 = 500$, where the center spike, sidelobes, and sub-sidelobes are marked with a triangle, squares, and circles, respectively; (c) Maximum sidelobes with respect to the distance $D$ with different RIS sizes and voxel sizes.}
    \label{fig:subpath}
\end{figure*}

\subsection{Subpath Correlations}

Eliminating the position index $t$ in \eqref{eq-y=Ax}, the single-view imaging problem is formulated as
\begin{equation}\label{eq-single-y=Ax}
\mathbf{y=Ax+z},
\end{equation}
where $\mathbf{x}$ is the sparse image vector, following a Bernoulli-Gaussian distribution with sparse rate $\alpha$, mean $\eta$, and covariance $\varsigma^2$.
$\mathbf{A}=\boldsymbol{\Omega}\mathbf{B}$ is the sensing matrix obtained by varying RIS phase shifts, which has enabled resolving various voxels.
Distinguishing different voxels with the RIS aperture is a challenging problem as the subpaths from the voxels to the RIS array are highly correlated.
Let $\mathbf{H}_{\rm{v,s}} = \left[\mathbf{h}_1^{\rm{v}}, \mathbf{h}_2^{\rm{v}}, \ldots, \mathbf{h}_N^{\rm{v}} \right]^{\rm{T}}$, where $\mathbf{h}_n^{\rm{v}}$ represents the subpath from the $n$-th voxel to the RIS.
The mutual correlation between the subpath from the $n_1$-th voxel to the RIS and that of the $n_2$-th voxel is given by
\begin{equation}\label{eq-subpath-corre}
\mu(n_1, n_2) = \frac{|\left<\mathbf{h}_{n_1}^{\rm{v}}, \mathbf{h}_{n_2}^{\rm{v}}\right>|}{\|\mathbf{h}_{n_1}^{\rm{v}}\|_2 \|\mathbf{h}_{n_2}^{\rm{v}}\|_2},
\end{equation}
where $n_1,n_2 \in [1, N]$.
The maximum sidelobe of $\mu(n_1, n_2)$ with a constant $n_1$ is defined as $\mathcal{F}_{\rm{v}}(n_1) = \operatorname{max}\left(\mu(n_1, n_2)\right)$, where $n_1\neq n_2$.
Consider the imaging scenario as shown in Fig. \ref{fig:subpath-a}, where the ROI is located in front of the RIS array at the distance $D$, and Table \ref{tab-case-study} enumerates the system parameters.
Taking $n_1=500$ as an example, the subpath correlation $\mu(n_1=500, n_2)$ is depicted in Fig. \ref{fig:subpath-b}, where several sidelobe peaks (marked with squares) are located around the center spike (marked with a triangle).
Moreover, the sidelobes are surrounded by sub-sidelobes (marked with circles).
The sidelobes and sub-sidelobes correspond to the voxels in the range and cross-range directions (\textbf{\textcolor[RGB]{107,166,61}{green}} and \textbf{\textcolor[RGB]{156,122,91}{brown}} voxels, respectively) with respect to the considered voxel (the \textbf{\textcolor[RGB]{0,113,188}{blue}} voxel) in Fig. \ref{fig:subpath-a}, respectively.
Typically, the highest sidelobe $\mathcal{F}_{\rm{v}}(n_1)$ in Fig. \ref{fig:subpath-b} poses challenges in resolving voxels.
$\mathcal{F}_{\rm{v}}(n_1)$ exhibits negative correlations with respect to voxel sizes and RIS aperture sizes, but is positively correlated to $D$, as illustrated in Fig. \ref{fig:subpath-c}.

\subsection{PSF of the Sensing Matrix}

In traditional SAR imaging, the PSF has been a crucial metric to evaluate imaging abilities \cite{gao2018efficient}, which is independent of the observed targets.
In this study, we adopt the PSF concept of CS theory, as proposed in \cite{lustig2007sparse}, to assess the effectiveness of the proposed imaging system. The PSF is given as
\begin{equation}\label{psf}
\operatorname{PSF}(n_1, n_2) = \frac{|\left<\mathbf{A}(:, n_1), \mathbf{A}(:, n_2)\right>|}{\|\mathbf{A}(:, n_1)\|_2 \|\mathbf{A}(:, n_2)\|_2},
\end{equation}
where $n_1,n_2 = 1, 2, \ldots, N$, and $\mathbf{A}(:, n_1)$ is the $n_1$-th column of $\mathbf{A}$.
Regarding the columns of $\mathbf{A}$ as the ``steering vectors'' of the corresponding voxels, the PSF in \eqref{psf} measures the mutual coherence of these steering vectors.
Ideally, these $N$ vectors should be orthogonal to each other, allowing for the resolution of various voxels with minimal uncertainty.
However, in practical systems, the steering vectors can be highly correlated, leading to aliasing artifacts in the reconstructed images.
Benefiting from the intelligent reconfiguration abilities of the RIS, its phase shifts can be optimized to minimize the PSF.
Equivalently, the mutual coherence should be minimized, given as \cite{hu2022metasketch,zhu2023ris,sankar2023coded}
\begin{equation} \label{eq-beta}
\begin{aligned}
\beta(\mathbf{A}) & =  \sum_{\substack{n_1,n_2\in[1,N], \\ n_1\neq n_2}} \operatorname{PSF}(n_1, n_2) = \|\tilde{\mathbf{A}}^{H}\tilde{\mathbf{A}}-\mathbf{I}\|^2_{\rm{F}},
\end{aligned}
\end{equation}
where $\tilde{\mathbf{A}}$ is the column-normalized version of $\mathbf{A}$, making optimizing \eqref{eq-beta} difficult.
The genetic algorithm is invoked in \cite{hu2022metasketch}, whereas gradient descent-based algorithms are used in \cite{zhu2023ris,sankar2023coded}.
Here, we refer to \cite{sankar2023coded} to replace $\tilde{\mathbf{A}}$ in \eqref{eq-beta} by ${\mathbf{A}}$ and formulate the problem of RIS phase shift optimization as
\begin{equation}
\begin{array}{lll}
(\text{P1}) & \underset{\boldsymbol{\Omega}}{\operatorname{min}} & \beta'(\mathbf{A})=\|{\mathbf{A}}^{\rm{H}}{\mathbf{A}}-\mathbf{I}\|_{\rm{F}}^2, \\
& \text {s.t. } & \left|{\Omega}_{k,m}\right|=1, \ k\in[1,K],\ m\in[1,M],
\end{array}
\end{equation}
where $\Omega_{k,m}$ is the $(k,m)$-th element of $\boldsymbol{\Omega}$.
Then, the Algorithm \ref{ag1} is employed to obtain sub-optimal solutions of (P1), where the gradient of $\beta'(\mathbf{A})$ is given as
\begin{equation}\label{eq-gradient}
\nabla_{\boldsymbol{\Omega}}\beta'(\mathbf{A}) = \boldsymbol{\Omega}\mathbf{B}(\mathbf{B}^{\rm{H}}\boldsymbol{\Omega}^{\rm{H}}\boldsymbol{\Omega}\mathbf{B}-\mathbf{I})\mathbf{B}^{\rm{H}},
\end{equation}
with the computational complexity of $O(KN^2)$.

\begin{figure}[t]
\begin{algorithm}[H]
\caption{RIS phase shift optimization algorithm \cite{sankar2023coded}.}
\label{ag1}
\begin{algorithmic}[1]

        \State $\mathbf{input}$: $\mathbf{B}$, step length $\tau$, maximum iteration number $I_{\rm{max}}$.

        \State Initialize $\boldsymbol{\Omega}^{(0)}\in\mathbb{C}^{K\times M}$ as a random matrix.

        \State  $\mathbf{for} \  i = 1 \ {\rm{to}} \  I_{\rm{max}} \ \mathbf{do}$

        \State \hspace{0.5cm} $\tilde{\boldsymbol{\Omega}}^{(i)} = {\boldsymbol{\Omega}}^{(i-1)} - \tau \nabla_{\boldsymbol{\Omega}}\beta'(\mathbf{A})$.

        \State \hspace{0.5cm} ${{\Omega}}^{(i)}_{k,m} = \tilde{{\Omega}}^{(i)}_{k,m} / |\tilde{{\Omega}}^{(i)}_{k,m}|$, obtaining ${\boldsymbol{\Omega}}^{(i)}$.

        \State $\mathbf{end\ for}$

        \State $\mathbf{output}$: the optimized RIS phase shift matrix $\boldsymbol{\Omega}^{(I_{\rm{max}})}$.

\end{algorithmic}
\end{algorithm}
\end{figure}

\subsection{Geometric Constraint on Imaging Abilities}

Although optimizing RIS phase shifts is anticipated to improve imaging accuracy, certain geometric constraints have limited the imaging abilities of the proposed system.
As indicated in \eqref{eq-large} and \eqref{eq-new-A}, the properties of $\mathbf{A}$ are closely linked to the characteristics of the UE-ROI-RIS-AP paths.
Based on the PSF definition in \eqref{psf}, the following theorem reveals that the imaging capabilities of the proposed systems are limited by the geometric subpath correlation from the ROI to the RIS.

\begin{theorem}
\label{theorem}
The PSF of $\mathbf{A}$ and the subpath correlation from the ROI to the RIS have the following relationship when the number of measurements ($K$) is large:
\begin{equation}\label{eq-psf-subpath-corre}
\operatorname{PSF}(n_1, n_2) \approx \mu(n_1, n_2),
\end{equation}
where $n_1,n_2 = 1, 2, \ldots, N$.
\end{theorem}
\emph{Proof}: See Appendix \ref{appendix-psf-subpath-corre}. \hfill $\blacksquare$

This theorem suggests that
as the number of RIS configurations increases, the PSF of $\mathbf{A}$ exhibits characteristics similar to the subpath correlation depicted in Fig. \ref{fig:subpath-b}.
Thus, the voxels in range directions feature highly correlated steering vectors, making them hard to be resolved\footnote{
The imaging accuracy may be enhanced by choosing individual voxel sizes in range and cross-range directions to match the distinct resolutions. However, this study only considers cuboid voxels with $\xi_{\rm{vx}} = \xi_{\rm{vy}} = \xi_{\rm{vz}}$ for simplicity.
}.
Therefore, the considered 3D imaging problem is more challenging than previous 2D imaging studies \cite{jiang2023near,hu2022metasketch,zhu2023ris,sankar2023coded}, and the geometric subpath correlation from the ROI to the RIS imposes constraints on imaging performances.
Moreover, we denote $\mathcal{F}_{\rm{p}}(n_1) = \operatorname{max}\left(\operatorname{PSF}(n_1, n_2)\right)$ with $n_1\neq n_2$, which serves as a metric for evaluating the severity of possible aliasing artifacts.
Theorem \ref{theorem} indicates that $\mathcal{F}_{\rm{p}}(n_1)$ follows the same trends as illustrated in Fig. \ref{fig:subpath-c}. Thus, a larger RIS array size and a smaller distance $D$, which both contribute to a larger RIS-subtended angle $\psi$, are anticipated to realize higher imaging performances.
Since the mainlobe $\mathcal{F}_{\rm{p}}(n_1)$ mainly originates from the voxels along the range direction, as shown in Fig. \ref{fig:subpath-b}, a large $\psi$ can also help range information retrieval.
Moreover, the imaging accuracy manages a tradeoff between $\psi$ and the voxel size $\xi_{\rm{v}}$.
Smaller angles necessitate larger voxel sizes for high-accuracy imaging, while larger angles allow for higher resolutions (i.e., smaller $\xi_{\rm{v}}$).

\begin{figure*}
\centering
\captionsetup{font=footnotesize}
\begin{subfigure}[b]{0.265\linewidth}
\centering
\includegraphics[width=\linewidth]{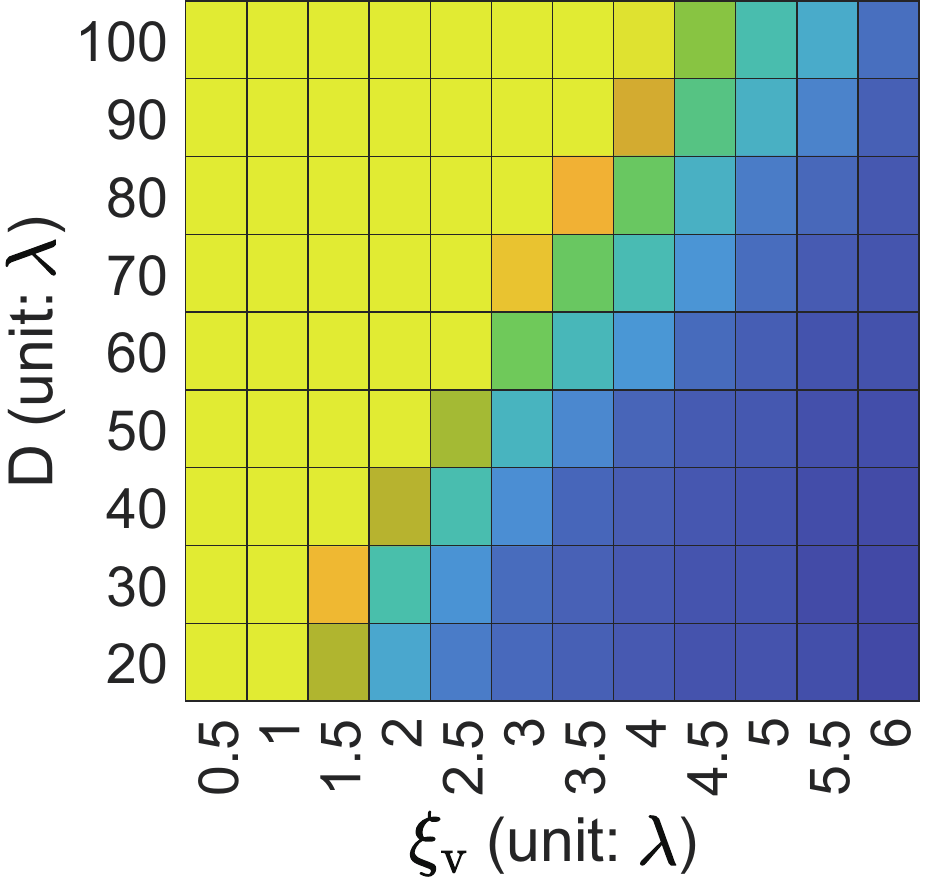}
\caption{SP \cite{dai2009subspace}}
\label{fig:single-1-a}
\end{subfigure}
\quad \quad
\begin{subfigure}[b]{0.243\linewidth}
\centering
\includegraphics[width=\linewidth]{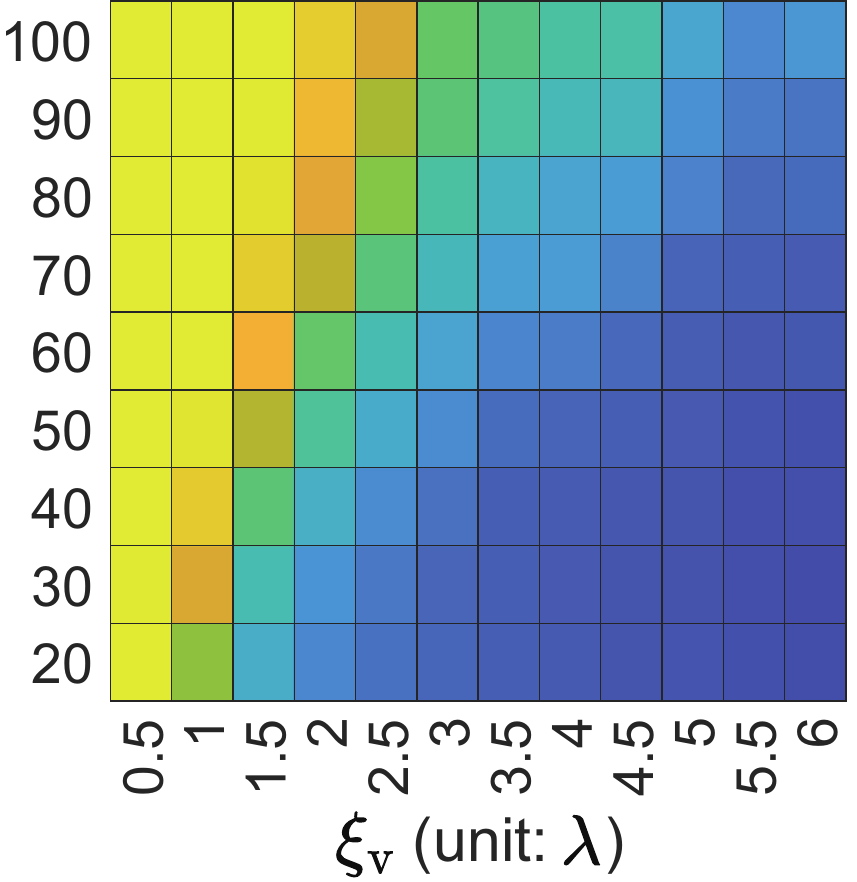}
\caption{GAMP \cite{rangan2011generalized}}
\label{fig:single-1-b}
\end{subfigure}
\quad \quad
\begin{subfigure}[b]{0.288\linewidth}
\centering
\includegraphics[width=\linewidth]{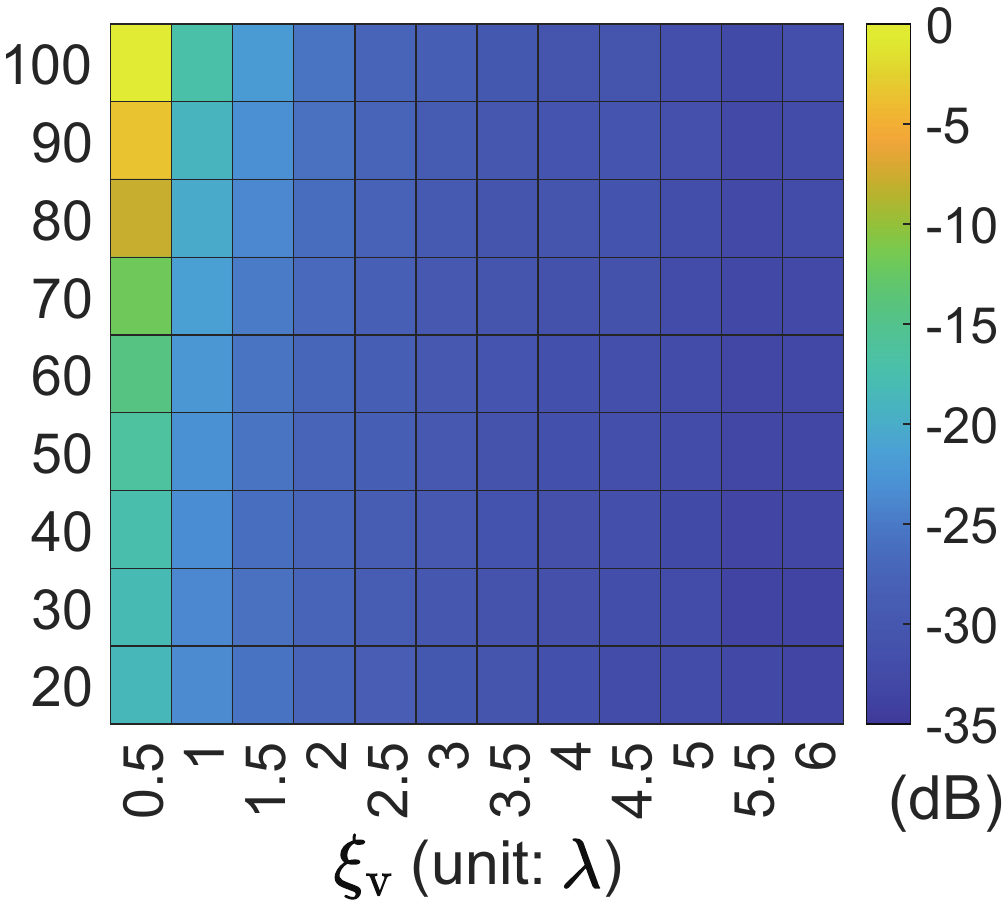}
\caption{SALS (lower bound)}
\label{fig:single-1-c}
\end{subfigure}
\caption{NMSE performances of different CS-based algorithms with various distances and voxel sizes.}
\label{fig:single-1}
\end{figure*}

\begin{remark}
Theorem \ref{theorem} is applicable to both randomly chosen continuous and discrete RIS phase shifts, provided that the phase shift values uniformly span the range $[0, 2\pi]$ and satisfy $\mathbb{E}\{e^{-j\omega_{k,m}}\} = 0$.
This means that RISs with discrete phase shifts can be employed in the proposed imaging schemes, as a ${b}$-bit RIS can generate $2^{{b}M}$ distinct radiation patterns, which are sufficient for capturing sparse target scattering features.
The imaging performances utilizing continuous and discrete RIS phase shifts are compared in Sec. \ref{subsubsec-multi-4} through simulations.
\end{remark}

\section{Numerical Results}
\label{sec-result}

\subsection{Performance Evaluation of Single-view 3D Imaging}
\label{subsec-result-single}

We begin by evaluating the performance of single-view imaging at individual UE positions. The simulation scenario is depicted in Fig. \ref{fig:model-a}, where the RIS array is positioned in the $yOz$ plane and centered at $[0, 0, 0]^{\rm{T}}$, with its dimensions aligned with the $y$- and $z$-axis.
The RIS element number and size correspond to the values provided in Table \ref{tab-case-study}.
The ROI is discretized into $N$ cuboid voxels with the side length $\xi_{\rm{v}}$.
We use an example frequency of $f = 3 \ \rm{GHz}$ in this subsection, presenting the results with electronic size $\lambda$, which can be extended to other frequency bands.
The evaluation of imaging performances involves assessing both the resolution and accuracy. These two aspects manage a tradeoff, as examined in Sec. \ref{subsubsec-sing-1}. In other parts of this section, the focus is on the assessment of imaging accuracy, assuming constant resolutions.
The image resolution is determined by balancing the desired accuracy, the angle subtended by the RIS $\psi$, the number of available measurements $K$, and computational and temporal resources.
Specifically, the image \textbf{resolution} is determined by the voxel sizes, while the imaging \textbf{accuracy} is assessed using the normalized mean square error (NMSE), given by
\begin{equation}\label{eq-nmse}
\operatorname{NMSE} = \frac{1}{I_{\rm{MC}}}\sum_{i = 1}^{I_{\rm{MC}}}\frac{\|\hat{\mathbf{x}}^{(i)} - \mathbf{x}\|^2_2}{\|\mathbf{x}\|^2_2},
\end{equation}
where $I_{\rm{MC}}=1000$ is the number of Monte Carlo simulations, and $\hat{\mathbf{x}}^{(i)}$ is the estimated image of $\mathbf{x}$ at the $i$-th simulation.

\subsubsection{Tradeoff between Image Resolution, Accuracy, and RIS-subtended Angle}
\label{subsubsec-sing-1}

We assess the single-frequency imaging performance at various distances $D$ (i.e., various $\psi$) and with distinct voxel sizes $\xi_{\rm{v}}$, considering a constant ROI size of $(20 \lambda)^3$.
We assume a sparse rate $\alpha = 4\%$ for a voxel size $\xi_{\rm{v}} = \lambda$, scaling proportionally as $\xi_{\rm{v}}$ changes.
We traverse $K=80$ random RIS phase shift configurations, with the UE and AP locations randomly generated within the space $\mathbb{G}=\{[x, y, z]^{\rm{T}}: 0 \leqslant x \leqslant 100, -50 \leqslant y \leqslant 50$, $-15 \leqslant z \leqslant 15 \ (\text{unit}: \lambda) \}$. The signal-to-noise ratio (SNR) of the measurements is set at $20\ \rm{dB}$.
The imaging performances of the SP, GAMP, and support aware least squares (SALS) algorithms are shown in Figs. \ref{fig:single-1-a}, \ref{fig:single-1-b}, and \ref{fig:single-1-c}, respectively.
SALS is the least squares algorithm with the awareness of sparse signal supports, representing the lower bound of other CS-based methods.
The results in Fig. \ref{fig:single-1} show a tradeoff between image resolution and accuracy as a function of the distance $D$ (i.e., angle $\psi$). With a larger voxel size $\xi_{\rm{v}}$ (i.e., lower image resolution), the image dimension $N$ is smaller, resulting in a lower sparse rate and reduced subpath correlations, as shown in Fig. \ref{fig:subpath-c}. This facilitates solving the inverse problem in \eqref{eq-single-y=Ax} with higher accuracy.
However, images constructed at longer distances have reduced resolutions or accuracies.
Fig. \ref{fig:single-1} reveals that the GAMP algorithm outperforms the SP algorithm in the considered scenarios, indicating that imaging performance is closely related to the sparse signal retrieving efficiency of CS-based methods. Lastly, the poor NMSE of SALS with $\xi_{\rm{v}} = \lambda / 2$ confirms the conclusion from Sec. \ref{sec-diffraction} that radio image resolutions should be larger than half the wavelength.

\begin{figure}
    \centering
    \includegraphics[width=0.8\linewidth]{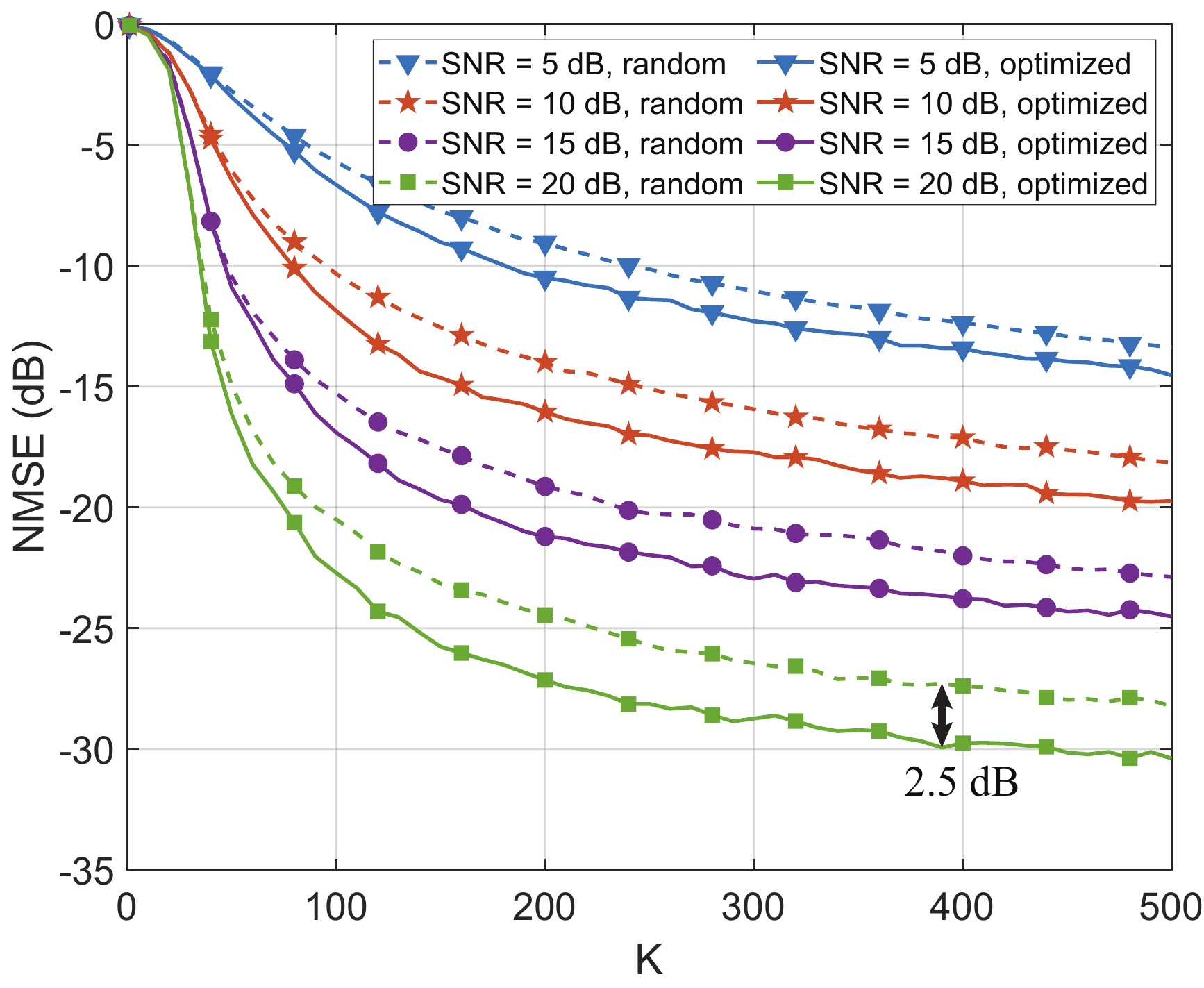}
    \captionsetup{font=footnotesize}
    \caption{NMSE performance with respect to the number $K$ of RIS phase shift configurations, with different SNRs and random or optimized RIS phase shifts.}
    \label{fig:single-2}
\end{figure}

\subsubsection{Performance Comparison with Random and Optimized RIS Phase Shifts}\label{subsubsec-sing-2}

We compare the imaging performances of random and optimized RIS phase shifts in single-view imaging using the GAMP algorithm.
The optimization Algorithm \ref{ag1} is employed with $I_{\rm{max}}=200$ and $\tau = 100$.
The measurement number $K$ is varied, and the influences of the SNR are discussed.
We set the UE and the AP to locate at $[30\lambda, 30\lambda, 10\lambda]^{\rm{T}}$ and $[20\lambda, 20\lambda, 30\lambda]^{\rm{T}}$, respectively, since the optimized phase shifts are functions of their locations.
The ROI with the size of $(20 \lambda)^3$ and located at $[40\lambda, 0, 0]^{\rm{T}}$ is divided into $10^3$ voxels, each with a size of $\xi_{\rm{v}} = 2\lambda$.
The sparse rate $\alpha=2\%$, and other simulation settings are the same as Sec. \ref{subsubsec-sing-1}.
The results, as shown in Fig. \ref{fig:single-2}, indicate that more measurements typically yield higher imaging accuracy.
Approximately 200 measurements with random phase shifts can diminish the NMSE to -25 dB at the SNR of 20 dB, leading to satisfactory imaging outcomes.
However, only 120 instances with optimized RIS phase shifts can achieve comparable results.
Furthermore, the NMSE gradually saturates when $K\geq 300$, supporting the conclusion from Theorem \ref{theorem} that imaging abilities are constrained by subpath correlations.
The optimized phase shifts enhance imaging accuracy by approximately 2.5 dB compared to random values when $K\geq 300$.
Furthermore, a higher SNR facilitates a lower NMSE, achieving saturation with a reduced number of measurements. Nonetheless, an increase in the number of RIS phase changes can mitigate high NMSE levels even at lower SNRs.

\subsection{Joint Multi-view 3D Imaging along a Continuous Trajectory}
\label{subsec-result-multi}

This subsection shifts the attention to evaluating the performance of joint multi-view imaging as proposed in Sec. \ref{sec-multi-view}.
The UE is assumed to move slowly along a continuous trajectory, which is randomly generated within the space $\mathbb{G}$.
Unless stated otherwise, the simulation parameters are listed in Table \ref{tab-multi}.
All other system settings remain the same as in Section \ref{subsec-result-single}.
Considering the large computational resources required to optimize RIS phase shifts at each UE position, we only utilize random values in this subsection.
To evaluate the performance of joint multi-view imaging, we employ the average normalized mean square error (aveNMSE) as the evaluation metric, which is calculated as follows:
\begin{equation}\label{eq-avenmse}
\operatorname{aveNMSE} = \frac{1}{I_{\rm{MC}}}\sum_{i = 1}^{I_{\rm{MC}}}  \frac{1}{T}\sum_{t=1}^T  \frac{\|\hat{\mathbf{x}}^{(i)}_t - \mathbf{x}_t\|^2_2}{\|\mathbf{x}_t\|^2_2},
\end{equation}
where $\hat{\mathbf{x}}^{(i)}_t$ is the estimated image at the $t$-th UE position in the $i$-th Monte Carlo simulation. We set $I_{\rm{MC}} = 1000$ in this subsection.

\begin{table}[t]
  \renewcommand{\arraystretch}{1.5}
  \centering
  \fontsize{8}{8}\selectfont
  \captionsetup{font=small}
  \captionof{table}{Simulation parameters of joint multi-view single-frequency 3D imaging.}\label{tab-multi}
  \begin{threeparttable}
    \begin{tabular}{c|c}
      \specialrule{1pt}{0pt}{-1pt}\xrowht{12pt}
      Parameters & Values \\
      \hline
      SNR & $20 \operatorname{dB}$ \\
      ROI size & $20\lambda \times 20\lambda \times 20\lambda$ \\
      AP position & $[2, 2, 3]^{\rm{T}}$ \\
      Imaging distance & $D = 50\lambda$ \\
      UE step length & $d_0 = 5\lambda$ \\
      Voxel size & $\xi_{\rm{v}} = 2\lambda$ \\
      Voxel number & $N = 10\times10\times10$ \\
      Sparse rate & $\alpha = 2\%$ \\
      RIS element size & $\xi_{\rm{s}} = \lambda / 2$ \\
      RIS element number & $M = 48 \times 48$ \\
      Reconfiguration number & $K = 80$ \\
      Mean of scattering coefficients & $\eta = 1$ \\
      Variance of scattering coefficient & $\varsigma^2 = 1$ \\
      Transition probabilities & $p_{01} = 0.1$ \\
      Temporal correlation & $\rho = 0.9$ \\
      \specialrule{1pt}{0pt}{0pt}
    \end{tabular}
  \end{threeparttable}
\end{table}

\subsubsection{Performance Comparison of Single-view and Joint Multi-view Imaging}

\begin{figure}[t]
\centering
\captionsetup{font=footnotesize}
\begin{subfigure}[b]{0.49\linewidth}
\centering
\includegraphics[width=\linewidth]{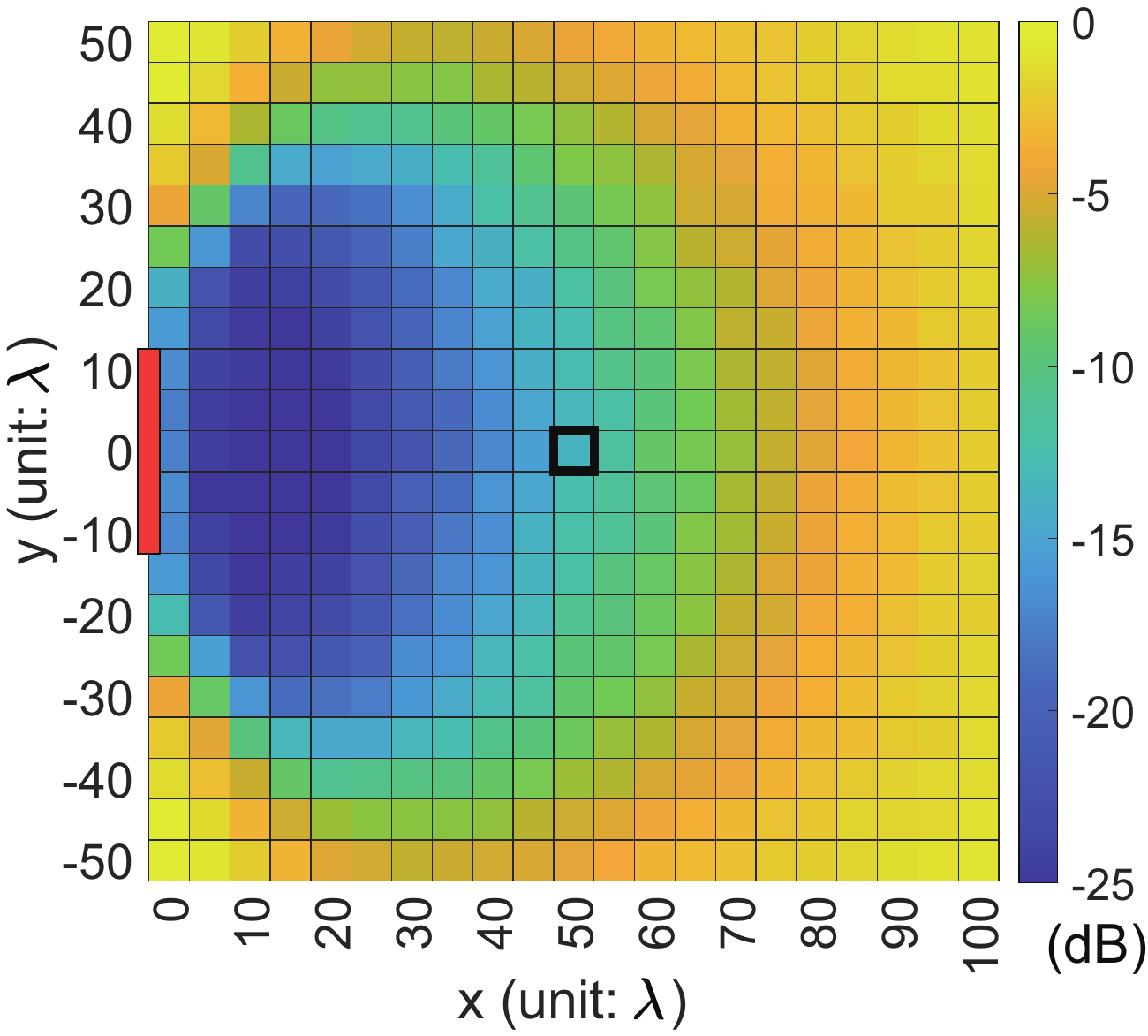}
\caption{}
\label{fig:multi-1-a}
\end{subfigure}
\begin{subfigure}[b]{0.49\linewidth}
\centering
\includegraphics[width=\linewidth]{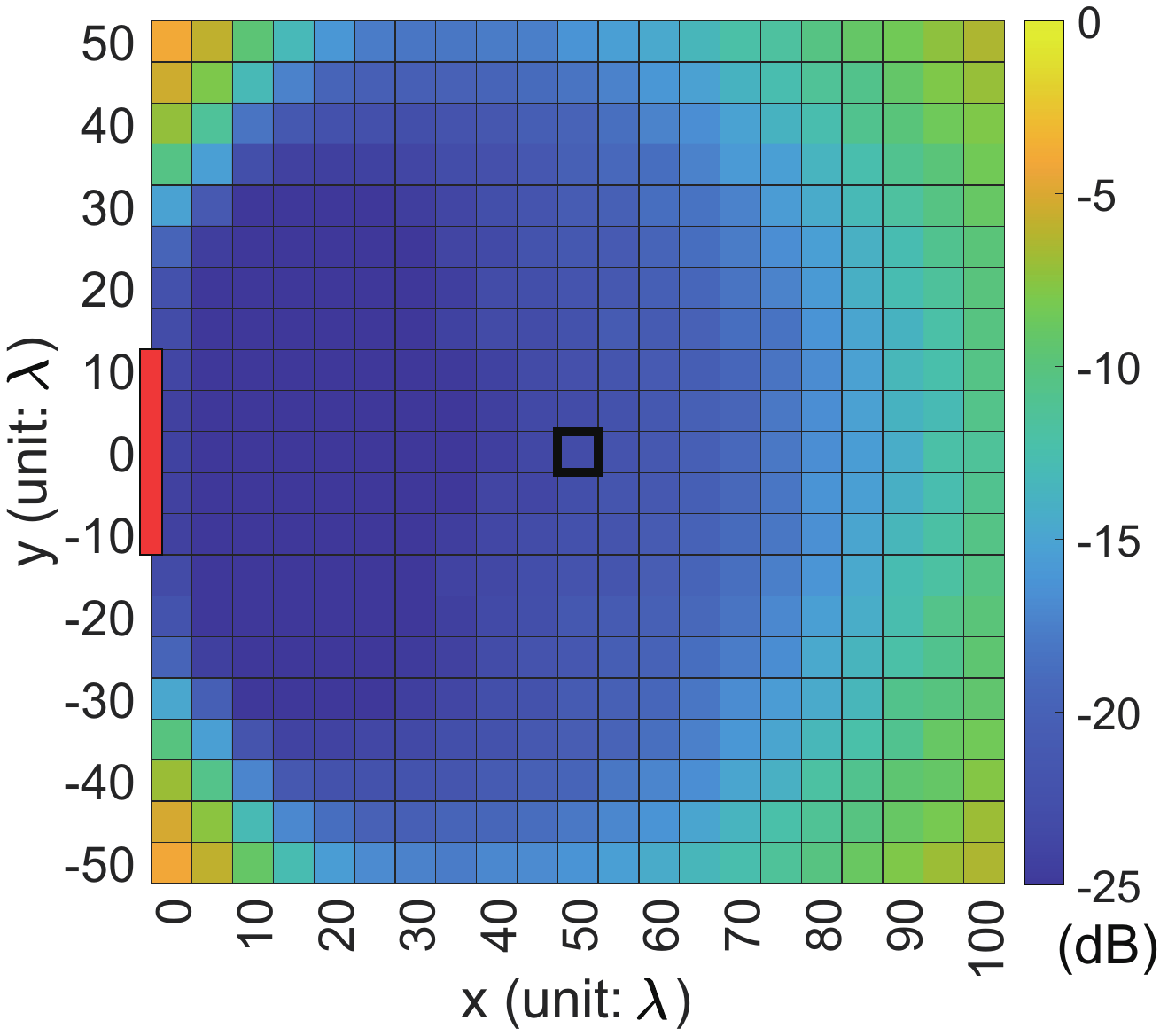}
\caption{}
\label{fig:multi-1-b}
\end{subfigure}

\begin{subfigure}[b]{0.45\linewidth}
\centering
\includegraphics[width=\linewidth]{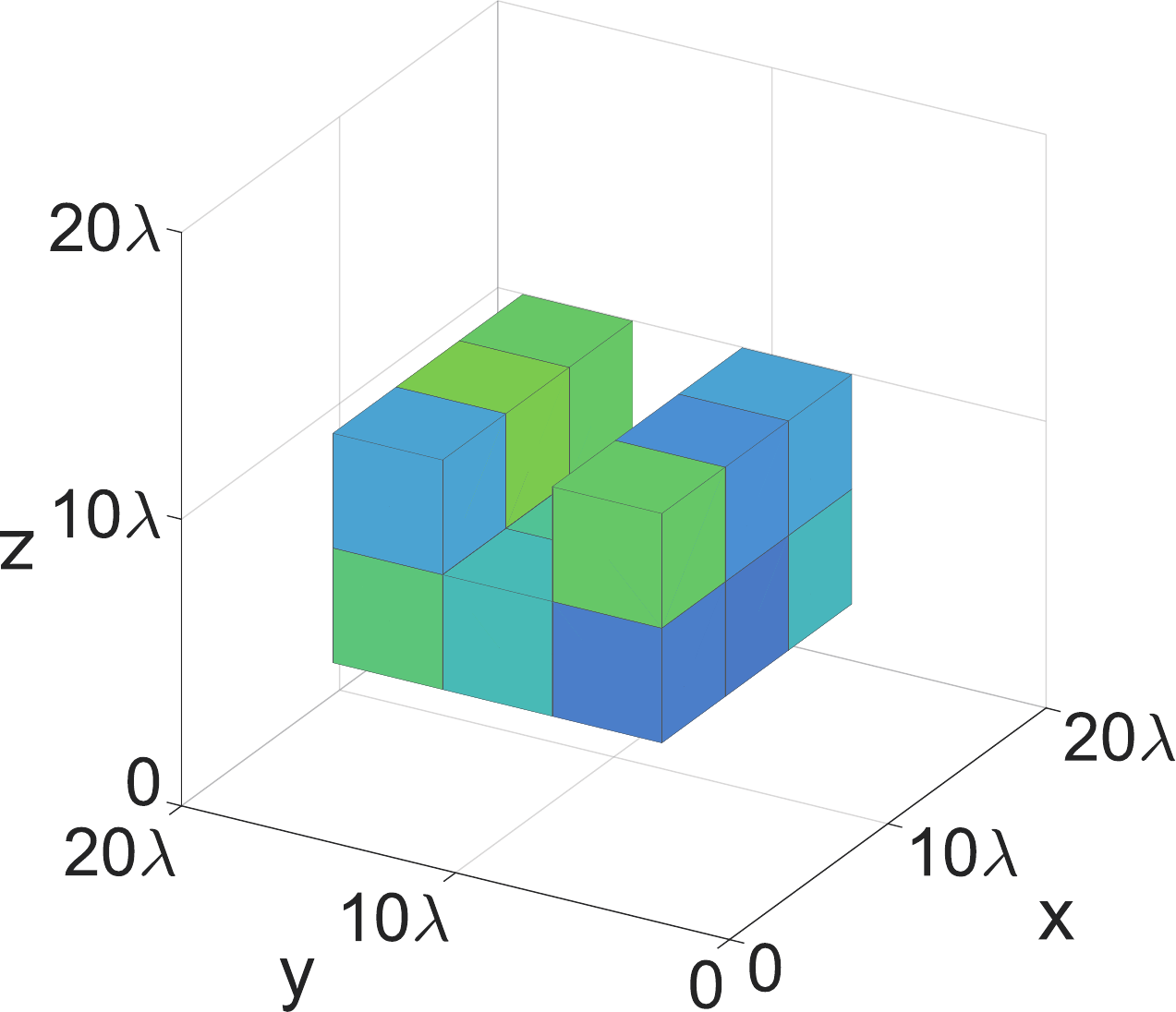}
\caption{}
\label{fig:multi-1-c}
\end{subfigure}
\begin{subfigure}[b]{0.45\linewidth}
\centering
\includegraphics[width=\linewidth]{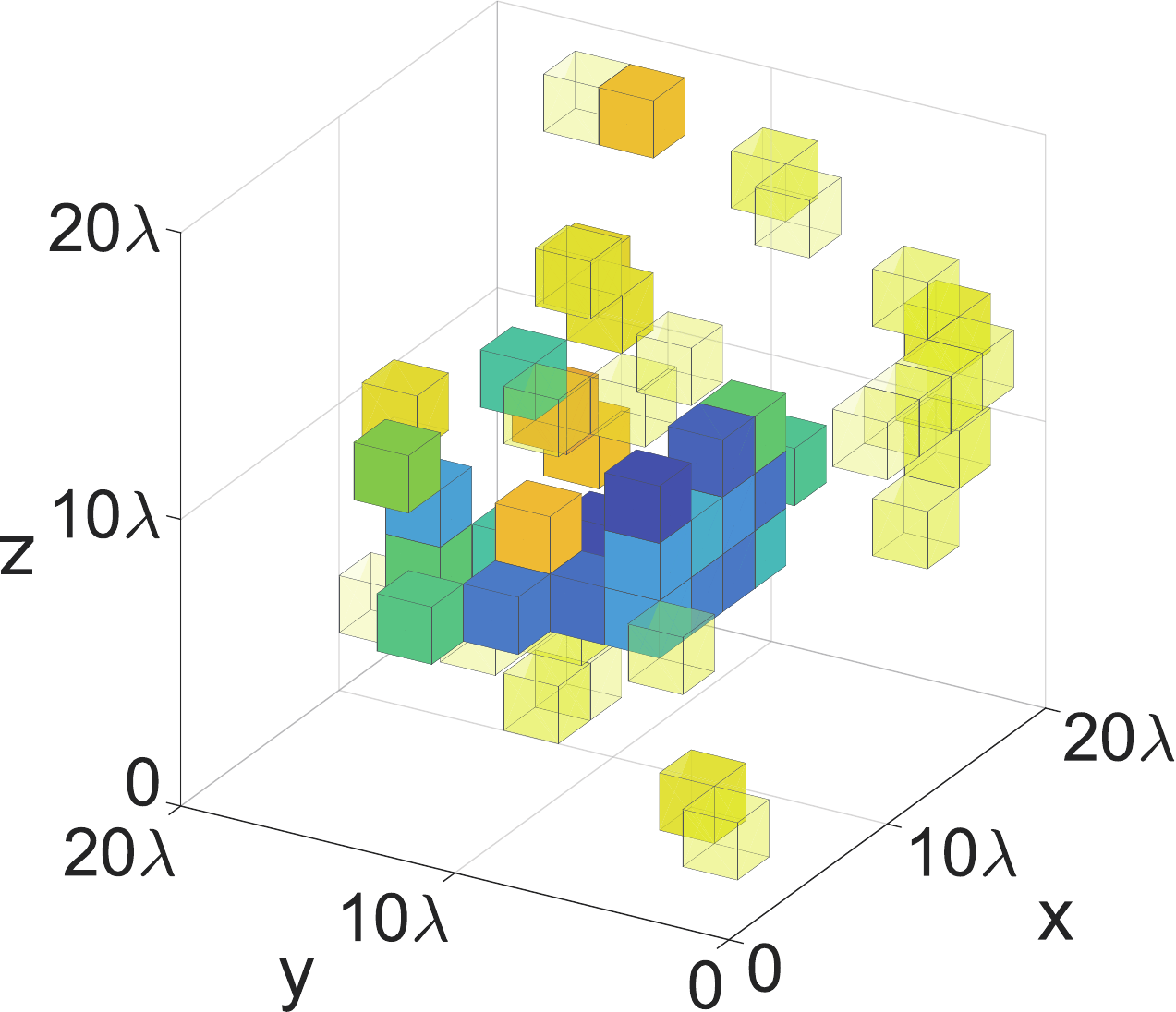}
\caption{}
\label{fig:multi-1-d}
\end{subfigure}
\begin{subfigure}[b]{0.7\linewidth}
\centering
\includegraphics[width=\linewidth]{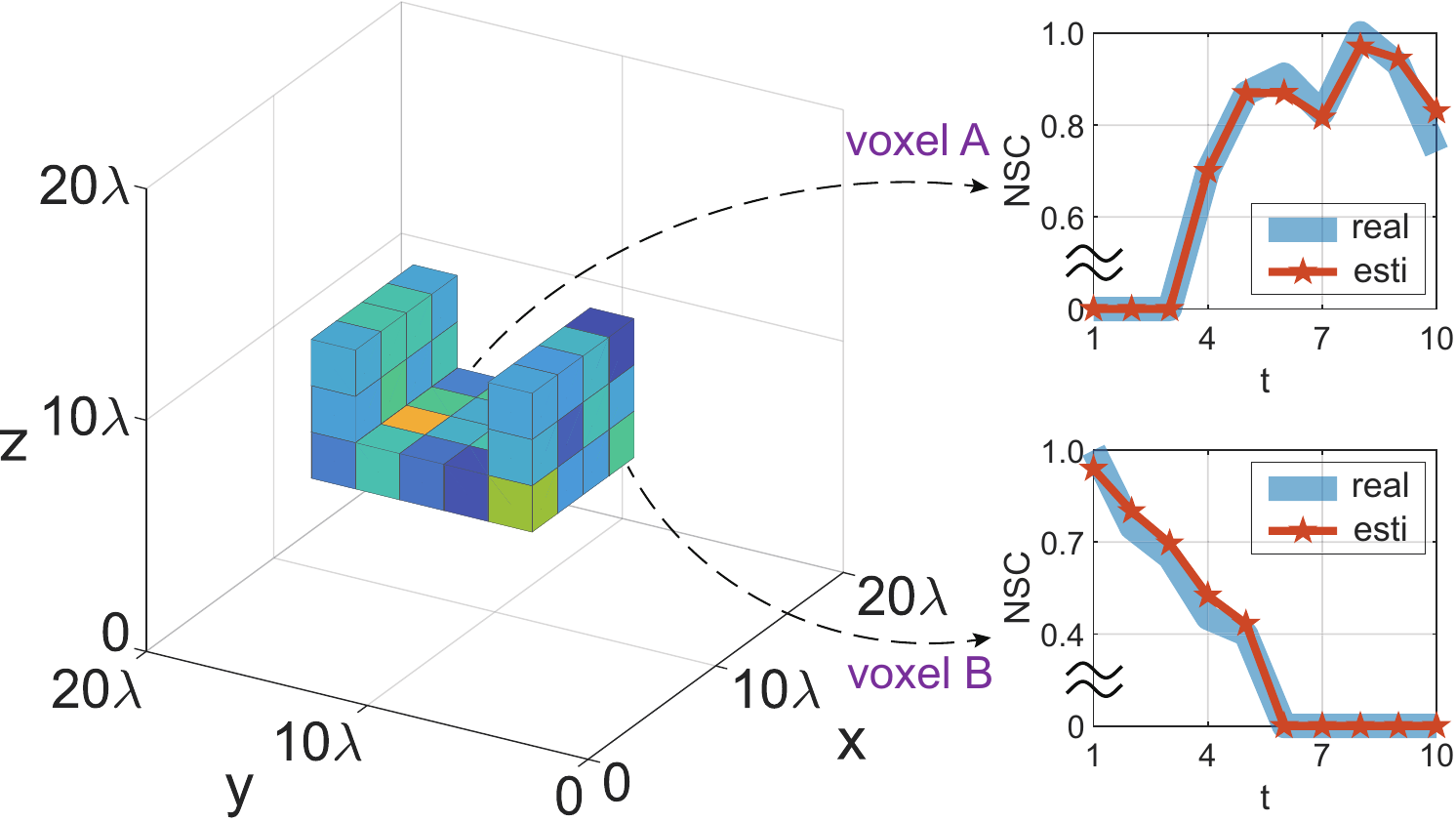}
\caption{}
\label{fig:multi-1-e}
\end{subfigure}
\caption{(a) aveNMSE performance of single-view imaging with various ROI positions; (b) aveNMSE performance of joint multi-view imaging with various ROI positions; (c) An intuitive example of the reconstructed images by single-view imaging with voxel size $\xi_{\rm{v}}=4\lambda$; (d) An intuitive example of the reconstructed images by single-view imaging with voxel size $\xi_{\rm{v}}=2\lambda$; (e) The left is an intuitive example of the reconstructed images by joint multi-view imaging with voxel size $\xi_{\rm{v}}=2\lambda$, and the right is the multi-view NSCs of voxels A and B (the $x$-axis represents the index of UE steps).}
\label{fig:multi-1}
\end{figure}

We compare the imaging capabilities of single-view and joint multi-view imaging by considering various center positions of the ROI in the space $\mathbb{G}$ with $z=0$.
For each ROI position and Monte Carlo simulation, a randomly generated continuous trajectory of the UE is used.
The imaging accuracies of single-view and joint multi-view imaging with $T = 10$ UE steps and $\xi_{\rm{v}} = 2\lambda$ are shown in Figs. \ref{fig:multi-1-a} and \ref{fig:multi-1-b}, respectively.
The red rectangular regions represent the RIS aperture from a bird's-eye view, and the accuracies of the reconstructed images are higher when the ROI is closer to the RIS, as the subpath correlations from the ROI to the RIS are relatively low.
Fig. \ref{fig:multi-1-b} demonstrates significant performance improvements in joint multi-view imaging, with larger imaging ranges and higher accuracies, compared to the results of single-view imaging shown in Fig. \ref{fig:multi-1-a}.
Examples of reconstructed 3D images from one of the multiple views are shown in Figs. \ref{fig:multi-1-c}, \ref{fig:multi-1-d}, and \ref{fig:multi-1-e} when the ROI is located at the bold black squares marked in Figs. \ref{fig:multi-1-a} and \ref{fig:multi-1-b}.
Different voxel colors represent different scattering coefficients, while scattering coefficients lower than 2\% of the maximum value are depicted as translucent yellow voxels.
Specifically, Figs. \ref{fig:multi-1-c} and \ref{fig:multi-1-d} showcase single-view imaging outcomes with varying voxel sizes.
Fig. \ref{fig:multi-1-c} delivers satisfactory results at the imaging resolution of $\xi_{\rm{v}}=4\lambda$. Conversely, Fig. \ref{fig:multi-1-d} exhibits several inaccurately detected voxels, failing at the imaging resolution of $\xi_{\rm{v}}=2\lambda$. In contrast, the proposed joint multi-view imaging algorithm successfully reconstructs the target's shape and scattering characteristics at $\xi_{\rm{v}}=2\lambda$, as depicted in Fig. \ref{fig:multi-1-e}. Hence, employing the proposed joint multi-view imaging algorithm facilitates an enhancement in imaging resolution.
Furthermore, the joint multi-view imaging results in the right part of Fig. \ref{fig:multi-1-e} provide an angular scattering response for each voxel with respect to the UE steps.
The true and estimated normalized scattering coefficients (NSCs) of voxels A and B demonstrate that voxel A can be observed from the fourth step onwards, while voxel B is only visible during the first five positions.
In conclusion, joint multi-view imaging outperforms single-view imaging in terms of imaging ranges, accuracies, resolutions, and anisotropy characterization.

\begin{figure}
    \centering
    \includegraphics[width=0.8\linewidth]{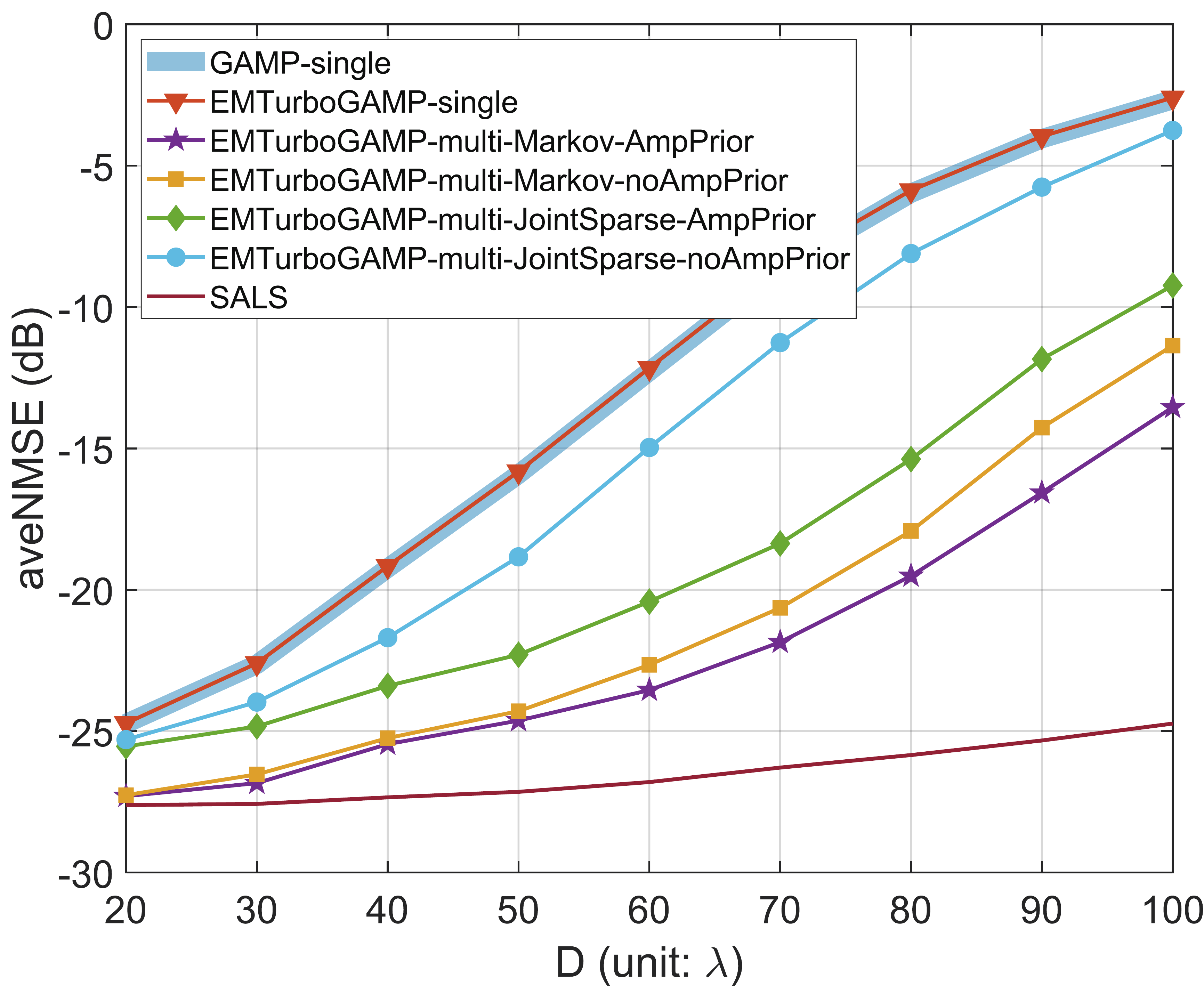}
    \captionsetup{font=footnotesize}
    \caption{aveNMSE performance comparison with different prior signal models in joint multi-view imaging with respect to the distance $D$, where ``single'' and ``multi'' represent single-view and joint multi-view imaging, respectively; ``Markov'' denotes the proposed binary Markov chain model, while ``JointSparse'' denotes the joint sparse prior signal model with constant supports among $\{\mathbf{x}_t\}_{t=1}^T$; ``AmpPrior'' and ``noAmpPrior'' imply employing the proposed Gauss-Markov amplitude prior signal model or not, respectively.}
    \label{fig:multi-2}
\end{figure}

\subsubsection{Performance Comparison of Various Prior Signal Models}

We utilize the EM-turbo-GAMP algorithm with different prior signal models to evaluate the effectiveness of the proposed joint multi-view imaging scheme, which simultaneously exploits slow signal support variation and smooth non-zero coefficient evolution.
Specifically, we compare the joint sparse prior signal model used in \cite{wu2021through}, which assumes the same signal support among $\{\mathbf{x}_t\}_{t=1}^T$, and the method that does not utilize the smooth coefficient evolution prior.
The integrated UE position number is set to $T = 10$.
The simulation results are shown in Fig. \ref{fig:multi-2}, where the aveNMSEs of various prior models increase as $D$ grows.
The comparable performances of GAMP and EM-turbo-GAMP in single-view imaging indicate that the EM-turbo-GAMP algorithm degrades to the GAMP algorithm when estimating a single image.
Various joint multi-view imaging prior models can enhance estimation accuracies, where the binary Markov chain prior signal support model outperforms the joint sparse prior model, and the smooth coefficient evolution prior enhances imaging performances.
Moreover, the best imaging performance is achieved with the proposed joint multi-view imaging scheme, but there is a significant gap between its performance and the lower bounds specified by SALS, particularly for large distances.
Therefore, there is room for designing and employing better joint multi-view imaging algorithms to improve imaging accuracy.

\subsubsection{Performance Gain of Joint Multi-view Imaging vs. Image Correlation Degrees}
\label{subsubsec-multi-3}

\begin{figure}
    \centering
    \includegraphics[width=0.8\linewidth]{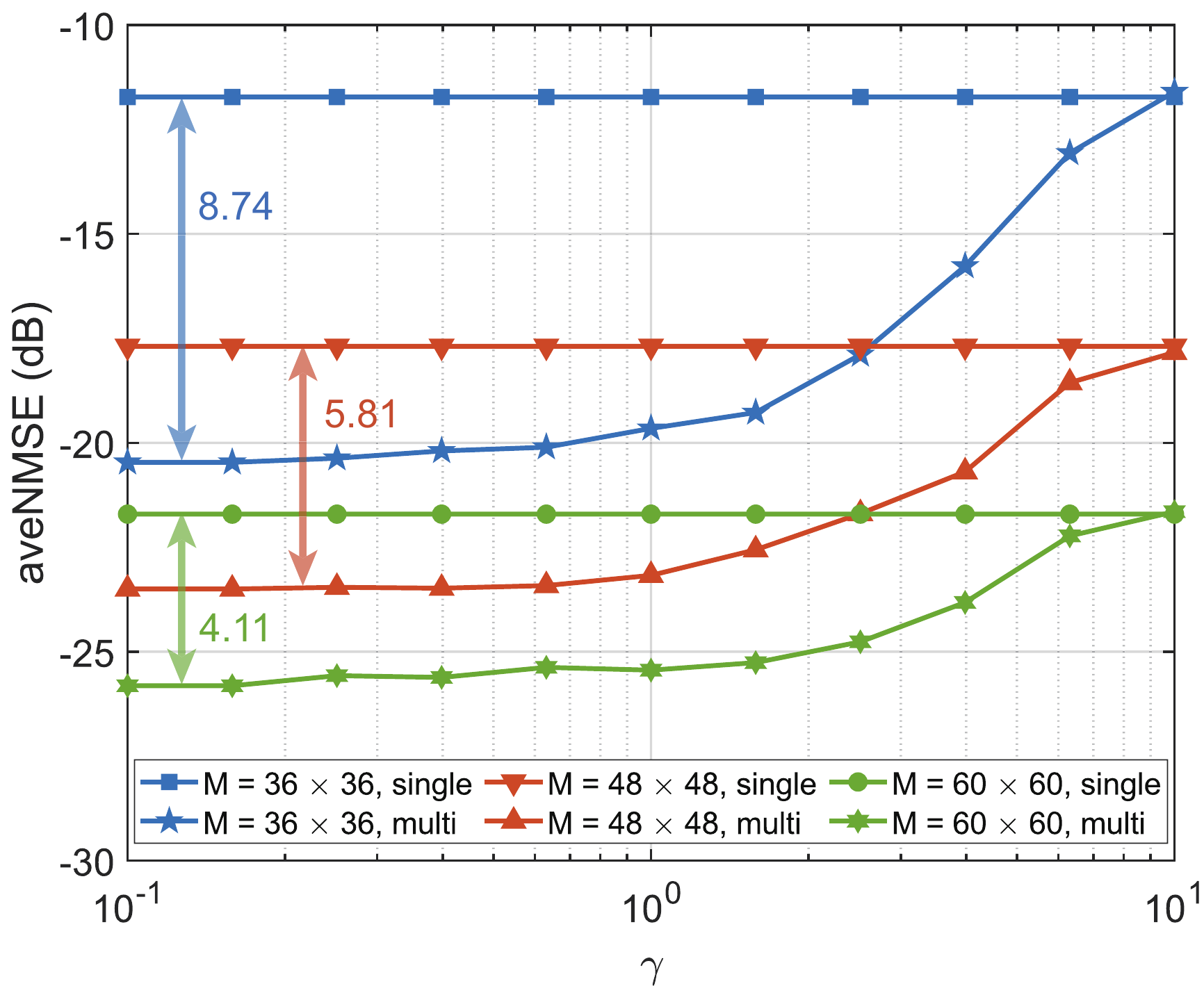}
    \captionsetup{font=footnotesize}
    \caption{NMSE performance comparison of joint multi-view imaging with various RIS sizes versus image correlation degrees, where ``single'' and ``multi'' represent single-view imaging and joint multi-view imaging, respectively.}
    \label{fig:multi-3}
\end{figure}

We investigate the performance gain of joint multi-view imaging compared to single-view imaging under different image correlation degrees, characterized by the Markov parameters $p_{01}$ and $\rho$ and related to the UE step interval $d_0$.
We set $d_0 = 2\lambda \gamma$, $p_{01} = 0.1\gamma$, and $\rho = 1 - 0.1\gamma$, where $\gamma \in [0, 10]$ is a scaling factor representing different correlation degrees.
The imaging accuracies of single-view and joint multi-view imaging are compared in Fig. \ref{fig:multi-3}, where the effects of RIS aperture sizes are also explored.
The aveNMSEs of single-view imaging remain constant as $\gamma$ varies since multi-view image correlations are not considered.
However, significant performance gains of joint multi-view imaging are observed when $\gamma < 1$, and the varying correlation degrees have little impact on image accuracies.
As $\gamma$ increases from 1 to 10, the aveNMSEs gradually increase and reach levels comparable to single-view imaging when $\gamma = 10$ (i.e., $p_{01} = 1$, and $\rho = 0$).
This suggests that the performance gains of joint multi-view imaging are inversely related to the correlation degree $\gamma$, i.e., the available information contained in multi-view images.
Multi-view imaging converges to single-view imaging when no correlations exist in multi-view images.
Additionally, as indicated by \eqref{eq-diffraction} and the simulation results, large RIS apertures yield high imaging accuracies but result in low performance gains for joint multi-view imaging.
This is because large RIS apertures approach the NMSE lower bound with a constant SNR, leaving small room for joint multi-view imaging to achieve performance gains.

\subsubsection{Imaging Performance vs. the RIS Reconfiguration Number}
\label{subsubsec-multi-4}

\begin{figure}
    \centering
    \includegraphics[width=0.8\linewidth]{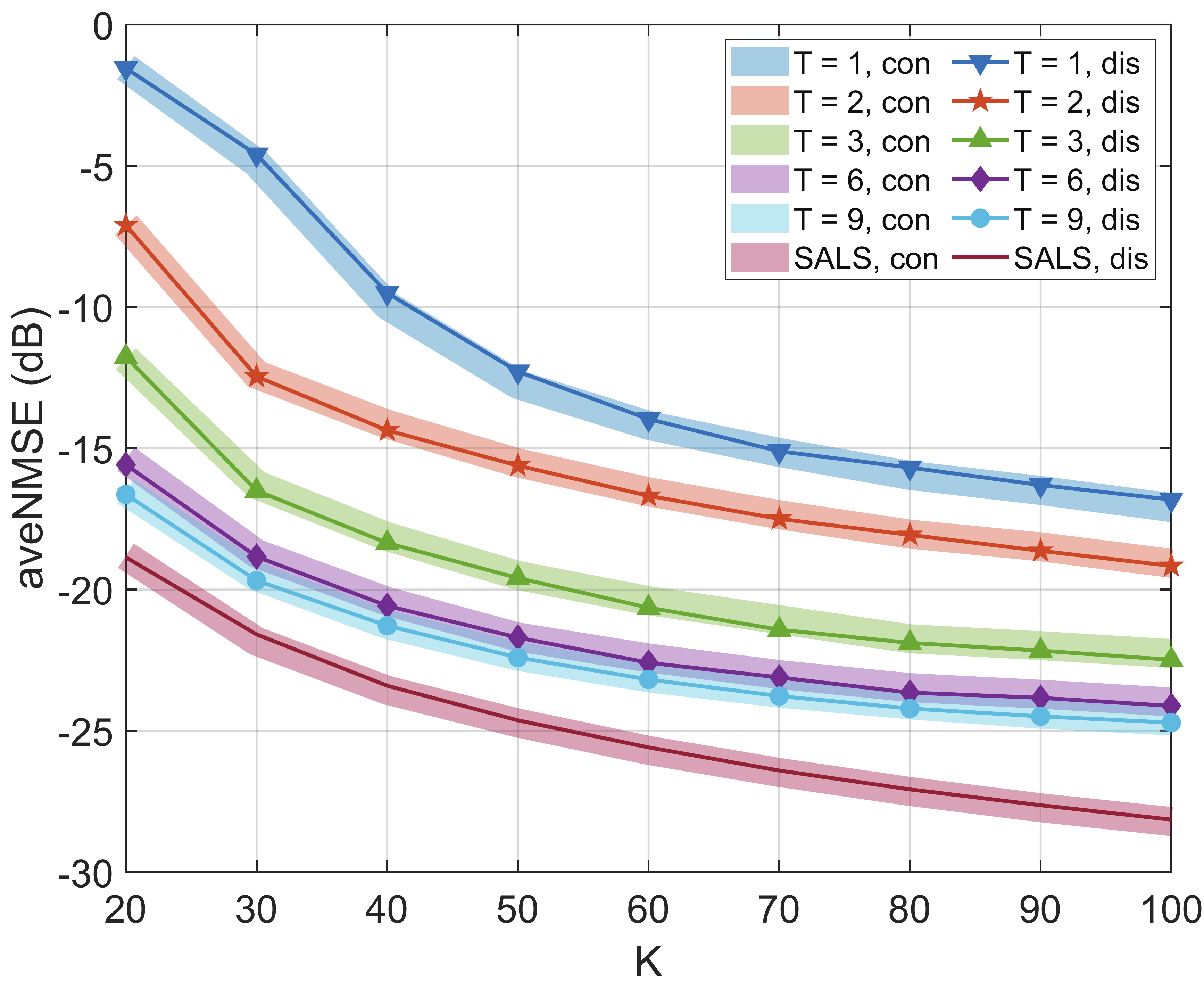}
    \captionsetup{font=footnotesize}
    \caption{NMSE performance comparison of single-view imaging and joint multi-view imaging with various UE steps versus the measurement number $K$ (``con'' and ``dis'' represent continuous and 1-bit discrete RIS phase shifts, respectively).}
    \label{fig:multi-4}
\end{figure}

We examine the required RIS reconfiguration numbers for single-view and joint multi-view imaging under different UE steps.
The simulation results are depicted in Fig. \ref{fig:multi-4}, where $T = 1$ corresponds to single-view imaging.
The aveNMSEs decrease as the number of RIS reconfigurations $K$ increases because a larger $K$ provides more measurements and captures more information about the scattering characteristics of the ROI.
Joint multi-view imaging reduces the number of required RIS configurations to achieve a certain imaging accuracy, enabling a low pilot overhead while maintaining satisfactory image quality.
Specifically, the aveNMSE of joint multi-view imaging combining $T = 9$ adjacent UE positions with $K = 20$ RIS reconfigurations achieves the same performance as single-view imaging with $K = 100$, resulting in an 80\% reduction in pilot symbols and occupied communication resources.
Consequently, the communication performance can be improved when the whole time resource is limited.
Furthermore, the performance gains of joint multi-view imaging are substantial when integrating the imaging processes of the first two or three adjacent positions, but diminish when $T > 6$.
Considering the increasing computing overhead of the EM-turbo-GAMP algorithm with $T$, the number of combined positions for joint multi-view imaging should be carefully selected to strike a balance between imaging accuracy and computing resources.
Lastly, the simulation results indicate that 1-bit random discrete RIS phase shifts can generate sufficient radiation patterns in the simulation scenarios, achieving comparable performances to random continuous phase shifts.

\subsubsection{Imaging Resolution Analysis}
\label{subsubsec-multi-5}

\begin{figure}
    \centering
    \includegraphics[width=0.8\linewidth]{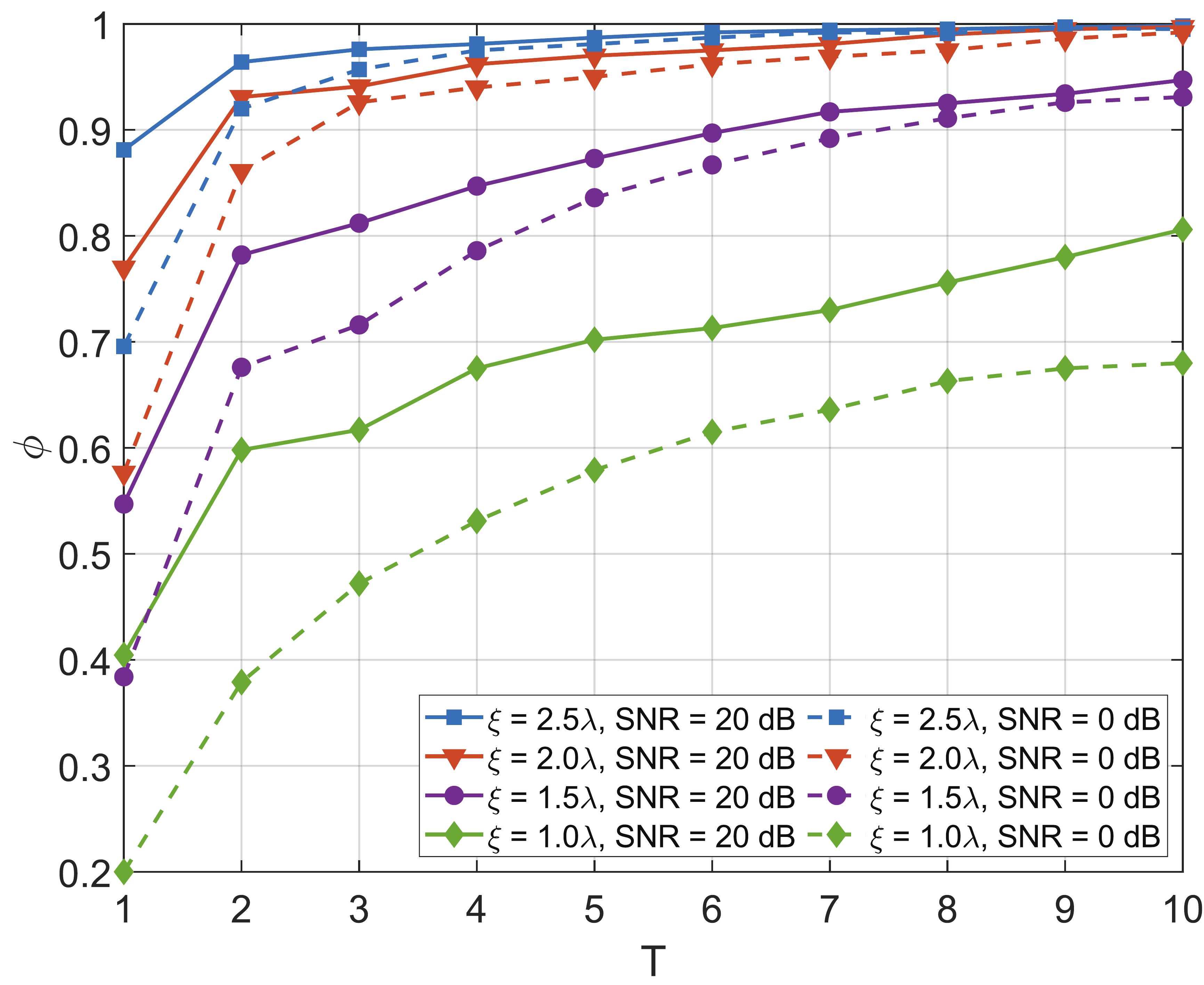}
    \captionsetup{font=footnotesize}
    \caption{Success detection rate with respect to the number of views for different voxel sizes and SNRs.}
    \label{fig:multi-5}
\end{figure}

We deploy two closely situated point targets to assess the enhancements in the achievable imaging resolution through our methodology. These targets are positioned within two adjacent voxels, with their separation corresponding to the voxel size $\xi_{\rm{v}}$. The RIS array, comprising $36\times36$ adjustable elements, is located at a distance of $D=60\lambda$ from the ROI.
Other configurations for the simulations align with the parameters detailed in Table \ref{tab-multi}.
Given that imaging outcomes might feature falsely detected voxels, as exemplified in Fig. \ref{fig:multi-1-c}, we define that the two targets are successfully detected solely when their corresponding voxels possess the highest scattering coefficients in the imaging results.
To evaluate imaging resolution, we introduce the success detection rate, $\phi$, as a metric. Drawing upon $I_{\rm{MC}}=1000$ Monte Carlo simulations, a $\phi$ exceeding the benchmark $\phi_0$ indicates that the imaging resolution, as defined by the voxel size $\xi_{\rm{v}}$, can be achieved with the proposed methods.
The outcomes, as illustrated in Fig. \ref{fig:multi-5}, show a noticeable increase in $\phi$ with an augmented number of views $T$, verifying the efficiency of the proposed joint multi-view imaging algorithm in enhancing imaging resolutions.
Setting $\phi_0=80\%$, we can achieve the imaging resolution of $2.5\lambda$ with single-view imaging at the SNR of 20 dB, $2\lambda$ with two different views, $1.5\lambda$ with three views, and $\lambda$ with ten or more views, as documented in Table \ref{tab-resolution}.
The success detection rate $\phi$ deteriorates as the SNR decreases from 20 dB to 0 dB. However, incorporating additional views in joint multi-view imaging can elevate $\phi$ and improve imaging resolution. Remarkably, the performance lines at the SNR of 0 dB can converge towards those at the SNR of 20 dB with an increased $T$, suggesting that multiple views can alleviate the adverse effects of low SNR on $\phi$.
Table \ref{tab-resolution} indicates that lower SNRs may achieve lower imaging resolutions. Consequently, enlarging the voxel size can compensate for this, enabling the derivation of reasonable imaging outcomes under low SNR conditions.

\begin{table}[t]
\renewcommand{\arraystretch}{1.5}
  \centering
  \fontsize{8}{8}\selectfont
  \captionsetup{font=small}
  \captionof{table}{The least number of views for different imaging resolutions and SNRs with the threshold $\phi_0=80\%$.}\label{tab-resolution}
  \begin{threeparttable}
    \begin{tabular}{c|c|c}
      \specialrule{1pt}{0pt}{-1pt}\xrowht{12pt}
      Resolution & SNR = 20 dB & SNR = 0 dB \\
      \hline
      $2.5\lambda$ & 1 & 2 \\
      $2.0\lambda$ & 2 & 2 \\
      $1.5\lambda$ & 3 & 5 \\
      $1.0\lambda$ & 10 & / \\
      \specialrule{1pt}{0pt}{0pt}
    \end{tabular}
  \end{threeparttable}
\end{table}

\section{Conclusion}
\label{sec:conclusion}
In this study, we present a novel RIS-aided 3D imaging scheme using joint multi-view image correlations and a single frequency. We underline the necessity of accounting for anisotropic scattering in multi-view imaging, develop models considering occlusion effects and anisotropic scattering as Markov processes, and use the EM-turbo-GAMP algorithm for image reconstruction. Our findings indicate that single-view imaging performance is limited by the geometric subpath correlation from the ROI to the RIS, but can be enhanced by optimizing RIS phase shifts.
Through simulations, we investigate and validate the tradeoff between the imaging resolution, accuracy, and RIS-subtended angle in single-view imaging, confirming the effectiveness of RIS phase shift optimization.
Additionally, the results show that our proposed joint multi-view imaging algorithm significantly improves imaging accuracy and reduces pilot overhead, especially when multi-view images possess high correlations.

\begin{appendices}
\section{}\label{appendix-psf-subpath-corre}

In this section, we present the proof of Theorem \ref{theorem}.
Using $\eqref{htk}$ and $\eqref{eq-y=Ax}$, we can give the $k$-th row of $\mathbf{A}$ by eliminating the UE position index $t$:
\begin{equation}
\begin{aligned}
\mathbf{A}(k, :) = g\mathbf{h}_{\rm{s,a}}^{\rm{T}}\operatorname{diag}(\boldsymbol{\omega}_k)\mathbf{H}_{\rm{v,s}}^{\rm{T}}\operatorname{diag}(\mathbf{h}_{\rm{u,v}}).
\end{aligned}
\end{equation}
As a result, we can express
\begin{equation}
\mathbf{A} = g\mathbf{D} \mathbf{H}_{\rm{v,s}}^{\rm{T}} \operatorname{diag}(\mathbf{h}_{\rm{u,v}}),
\end{equation}
where 
\begin{equation}
\mathbf{D} = \left[\mathbf{h}_{\rm{s,a}}, \ldots, \mathbf{h}_{\rm{s,a}}\right]^{\rm{T}} \odot \boldsymbol{\Omega}.
\end{equation}
Thus, the PSF of $\mathbf{A}$ can be reformulated as
\begin{equation}\label{psf-appendix}
\begin{aligned}
\operatorname{PSF}(n_1,n_2) & = \frac{\left|\left(\mathbf{D h}_{n_1}^{\rm{v}}\right)^{\rm{H}}\left(\mathbf{D h}_{n_2}^{\rm{v}}\right)\right|}{\left\|\left(\mathbf{D h}_{n_1}^{\rm{v}}\right)\right\|_{2}\left\|\left(\mathbf{D h}_{n_2}^{\rm{v}}\right)\right\|_{2}},
\end{aligned}
\end{equation}
where $\mathbf{A}(:, n) = g{h}_{{\rm{u,v}}, n}\left(\mathbf{D h}_{n}^{\rm{v}}\right)$, and ${h}_{{\rm{u,v}}, n}$ is the $n$-th element of $\mathbf{h}_{{\rm{u,v}}}$.
Denote $\mathbf{G}=\mathbf{D}^{\rm{H}} \mathbf{D} \in \mathbb{C}^{M \times M}$, and the $m$-th row of $\mathbf{G}$ is given as
\begin{equation}\label{0422-1}
\begin{aligned}
\mathbf{G}(m,:)= {h}_{{\rm{s,a}},m}^{*} \mathbf{h}_{{\rm{s,a}}}^{\rm{T}} \odot (\boldsymbol{\Omega}(:, m)^{\rm{H}}\boldsymbol{\Omega}),
\end{aligned}
\end{equation}
where ${h}_{{\rm{s,a}},m}$ is the $m$-th element of $\mathbf{h}_{{\rm{s,a}}}$.
The first term of the Hadamard product in \eqref{0422-1} is related to the channel from the RIS to the AP, and the second term represents the matching of the columns of $\boldsymbol{\Omega}$.
Considering that RIS element phase shifts are chosen randomly in the range $[0, 2\pi]$, the element $\Omega_{k,m} = e^{-j\omega_{k,m}}$ in $\boldsymbol{\Omega}$ follows a uniform distribution with zero mean.
Thus, we have $\mathbb{E}\{\Omega_{k, m_1}^{*} \Omega_{k, m_2}\} = 1$ when $m_1 = m_2$, and $\mathbb{E}\{\Omega_{k, m_1}^{*} \Omega_{k, m_2}\} = 0$ when $m_1 \neq m_2$.
Since
\begin{equation}
\boldsymbol{\Omega}(:, m_1)^{\rm{H}} \boldsymbol{\Omega}(:, m_2)=\Omega_{1, m_1}^{*} \Omega_{1, m_2} + \cdots + \Omega_{K, m_1}^{*} \Omega_{K, m_2},
\end{equation}
we have $\mathbb{E}\{\boldsymbol{\Omega}(:, m_1)^{\rm{H}} \boldsymbol{\Omega}(:, m_2)\} = K$ when $m_1 = m_2$ with aligned element phases, and $\mathbb{E}\{\boldsymbol{\Omega}(:, m_1)^{\rm{H}} \boldsymbol{\Omega}(:, m_2)\} = 0$ when $m_1 \neq m_2$ and $K \to \infty$, since the element phases are not aligned.
Therefore, the $m$-th element in the $m$-th row of $\mathbf{G}$ has the largest modulus, $|{h}_{{\rm{s,a}},m}|^2 K$, whereas other elements approach 0 as $K$ increases.
Assuming $|{h}_{{\rm{s,a}},m}| \approx |{h}_{{\rm{s,a}},0}|$ and ignoring the correlations between the rows of $\mathbf{G}$, we have $\mathbf{G}=\mathbf{D}^{\rm{H}} \mathbf{D} \approx |{h}_{{\rm{s,a}},0}|^2 K\mathbf{I}$, where $\mathbf{I}$ denotes an identity matrix.
Thus, $\mathbf{D}$ exhibits the properties of a unitary matrix with inner product invariance, leading to
\begin{equation}\label{eq-xxx}
\frac{\left|\left(\mathbf{D h}_{n_1}^{\rm{v}}\right)^{\rm{H}}\left(\mathbf{D h}_{n_2}^{\rm{v}}\right)\right|}{\left\|\left(\mathbf{D h}_{n_1}^{\rm{v}}\right)\right\|_{2}\left\|\left(\mathbf{D h}_{n_2}^{\rm{v}}\right)\right\|_{2}} \approx \frac{|\mathbf{h}_{n_1}^{{\rm{v}}\rm{H}} \mathbf{h}_{n_2}^{\rm{v}}|}{\left\|\mathbf{h}_{n_1}^{\rm{v}}\right\|_{2}\left\|\mathbf{h}_{n_2}^{\rm{v}}\right\|_{2}}.
\end{equation}
By substituting \eqref{eq-xxx} into \eqref{eq-subpath-corre} and \eqref{psf-appendix}, Theorem \ref{theorem} is proven.
Note that the approximation shown in \eqref{eq-xxx} typically cannot attain equality, but it approaches equality as $K$ increases.
\hfill $\blacksquare$
\end{appendices}

\bibliographystyle{IEEEtran}
\bibliography{ref}{}

\begin{thebibliography}{10}
\providecommand{\url}[1]{#1}
\csname url@samestyle\endcsname
\providecommand{\newblock}{\relax}
\providecommand{\bibinfo}[2]{#2}
\providecommand{\BIBentrySTDinterwordspacing}{\spaceskip=0pt\relax}
\providecommand{\BIBentryALTinterwordstretchfactor}{4}
\providecommand{\BIBentryALTinterwordspacing}{\spaceskip=\fontdimen2\font plus
\BIBentryALTinterwordstretchfactor\fontdimen3\font minus
  \fontdimen4\font\relax}
\providecommand{\BIBforeignlanguage}[2]{{%
\expandafter\ifx\csname l@#1\endcsname\relax
\typeout{** WARNING: IEEEtran.bst: No hyphenation pattern has been}%
\typeout{** loaded for the language `#1'. Using the pattern for}%
\typeout{** the default language instead.}%
\else
\language=\csname l@#1\endcsname
\fi
#2}}
\providecommand{\BIBdecl}{\relax}
\BIBdecl

\bibitem{liu2022integrated}
F.~Liu, Y.~Cui, C.~Masouros, J.~Xu, T.~X. Han, Y.~C. Eldar, and S.~Buzzi,
  ``Integrated sensing and communications: Towards dual-functional wireless
  networks for 6{G} and beyond,'' \emph{IEEE J. Sel. Areas Commun.}, vol.~40,
  no.~6, pp. 1728--1767, Jun. 2022.

\bibitem{yang2021enabling}
J.~Yang, C.-K. Wen, S.~Jin, and X.~Li, ``Enabling plug-and-play and
  crowdsourcing {SLAM} in wireless communication systems,'' \emph{IEEE Trans.
  Wireless Commun.}, vol.~21, no.~3, pp. 1453--1468, Mar. 2022.

\bibitem{imani2020review}
M.~F. Imani \emph{et~al.}, ``Review of metasurface antennas for computational
  microwave imaging,'' \emph{IEEE Trans. Antennas Propag.}, vol.~68, no.~3, pp.
  1860--1875, Mar. 2020.

\bibitem{wei2022toward}
Z.~Wei, F.~Liu, C.~Masouros, N.~Su, and A.~P. Petropulu, ``Toward
  multi-functional 6{G} wireless networks: Integrating sensing, communication,
  and security,'' \emph{IEEE Commun. Mag.}, vol.~60, no.~4, pp. 65--71, Apr.
  2022.

\bibitem{sheen2001three}
D.~M. Sheen, D.~L. McMakin, and T.~E. Hall, ``Three-dimensional millimeter-wave
  imaging for concealed weapon detection,'' \emph{IEEE Trans. Microwave Theory
  Tech.}, vol.~49, no.~9, pp. 1581--1592, Sep. 2001.

\bibitem{lien20175g}
S.-Y. Lien, S.-L. Shieh, Y.~Huang, B.~Su, Y.-L. Hsu, and H.-Y. Wei, ``5{G} new
  radio: Waveform, frame structure, multiple access, and initial access,''
  \emph{IEEE Commun. Mag.}, vol.~55, no.~6, pp. 64--71, Jun. 2017.

\bibitem{tang2021path}
W.~Tang \emph{et~al.}, ``Path loss modeling and measurements for reconfigurable
  intelligent surfaces in the millimeter-wave frequency band,'' \emph{IEEE
  Trans. Commun.}, vol.~70, no.~9, pp. 6259--6276, Sep. 2022.

\bibitem{rahal2021ris}
M.~Rahal, B.~Denis, K.~Keykhosravi, B.~Uguen, and H.~Wymeersch, ``{RIS}-enabled
  localization continuity under near-field conditions,'' in \emph{Proc. IEEE
  22nd Int. Workshop Signal Process. Adv. Wireless Commun.}, Sep. 2021, pp.
  436--440.

\bibitem{chen2023multi}
W.~Chen, C.-K. Wen, X.~Li, and S.~Jin, ``Multi-timescale channel customization
  for transmission design in {RIS}-assisted {MIMO} systems,'' \emph{IEEE J.
  Sel. Areas Commun.}, vol.~41, no.~8, pp. 2397--2413, Aug. 2023.

\bibitem{castaldi2021joint}
G.~Castaldi, L.~Zhang, M.~Moccia, A.~Y. Hathaway, W.~X. Tang, T.~J. Cui, and
  V.~Galdi, ``Joint multi-frequency beam shaping and steering via
  space-time-coding digital metasurfaces,'' \emph{Adv. Funct. Mater.}, vol.~31,
  no.~6, Feb. 2021, Art. no. 2007620.

\bibitem{taha2022reconfigurable}
A.~Taha, H.~Luo, and A.~Alkhateeb, ``Reconfigurable intelligent surface aided
  wireless sensing for scene depth estimation,'' in \emph{Proc. IEEE ICC}, May
  2023, pp. 491--497.

\bibitem{jiang2023near}
Y.~Jiang, F.~Gao, Y.~Liu, S.~Jin, and T.~Cui, ``Near field computational
  imaging with {RIS} generated virtual masks,'' [Online]. Available:
  https://arxiv.org/abs/2304.11510.

\bibitem{ccetin2014sparsity}
M.~{\c{C}}etin \emph{et~al.}, ``Sparsity-driven synthetic aperture radar
  imaging: Reconstruction, autofocusing, moving targets, and compressed
  sensing,'' \emph{IEEE Signal Process. Mag.}, vol.~31, no.~4, pp. 27--40, Jul.
  2014.

\bibitem{dai2009subspace}
W.~Dai and O.~Milenkovic, ``Subspace pursuit for compressive sensing signal
  reconstruction,'' \emph{IEEE Trans. Inf. Theory}, vol.~55, no.~5, pp.
  2230--2249, May 2009.

\bibitem{rangan2011generalized}
S.~Rangan, ``Generalized approximate message passing for estimation with random
  linear mixing,'' in \emph{Proc. IEEE Int. Symp. Inf. Theory}, Aug. 2011, pp.
  2168--2172.

\bibitem{hu2022metasketch}
J.~Hu, H.~Zhang, K.~Bian, Z.~Han, H.~V. Poor, and L.~Song, ``Metasketch:
  Wireless semantic segmentation by reconfigurable intelligent surfaces,''
  \emph{IEEE Trans. Wireless Commun.}, vol.~21, no.~8, pp. 5916--5929, Aug.
  2022.

\bibitem{sankar2023coded}
R.~Sankar and S.~P. Chepuri, ``Coded aperture radar imaging using
  reconfigurable intelligent surfaces,'' in \emph{Proc. IEEE CAMSAP}, Dec.
  2023, pp. 171--175.

\bibitem{zhu2023ris}
S.~Zhu, Z.~Yu, Q.~Guo, J.~Ding, Q.~Cheng, and T.~J. Cui, ``{RIS}-assisted joint
  uplink communication and imaging: Phase optimization and {B}ayesian echo
  decoupling,'' [Online]. Available: https://arxiv.org/abs/2301.03817.

\bibitem{gao2018efficient}
J.~Gao, B.~Deng, Y.~Qin, H.~Wang, and X.~Li, ``An efficient algorithm for
  {MIMO} cylindrical millimeter-wave holographic 3-{D} imaging,'' \emph{IEEE
  Trans. Microwave Theory Tech.}, vol.~66, no.~11, pp. 5065--5074, Nov. 2018.

\bibitem{lustig2007sparse}
M.~Lustig, D.~Donoho, and J.~M. Pauly, ``Sparse {MRI}: The application of
  compressed sensing for rapid {MR} imaging,'' \emph{Magn. Reson. Med.},
  vol.~58, no.~6, pp. 1182--1195, Nov. 2007.

\bibitem{fromenteze2017single}
T.~Fromenteze, M.~Boyarsky, J.~Gollub, T.~Sleasman, M.~Imani, and D.~R. Smith,
  ``Single-frequency near-field {MIMO} imaging,'' in \emph{Proc. 11th Eur.
  Conf. Antennas Propag. (EUCAP)}, Mar. 2017, pp. 1415--1418.

\bibitem{tong2021joint}
X.~Tong, Z.~Zhang, J.~Wang, C.~Huang, and M.~Debbah, ``Joint multi-user
  communication and sensing exploiting both signal and environment sparsity,''
  \emph{IEEE J. Sel. Top. Signal Process.}, vol.~15, no.~6, pp. 1409--1422,
  Nov. 2021.

\bibitem{tong2022environment}
X.~Tong, Z.~Zhang, Y.~Zhang, Z.~Yang, C.~Huang, K.-K. Wong, and M.~Debbah,
  ``Environment sensing considering the occlusion effect: A multi-view
  approach,'' \emph{IEEE Trans. Signal Process.}, vol.~70, pp. 3598--3615, Jun.
  2022.

\bibitem{moses2004wide}
R.~L. Moses, L.~C. Potter, and M.~Cetin, ``Wide angle {SAR} imaging,'' in
  \emph{Proc. SPIE Algorithms Synth. Aperture Radar Imagery XI}, vol. 5427,
  Sep. 2004, pp. 164--175.

\bibitem{huang2023joint}
Y.~Huang, J.~Yang, W.~Tang, C.-K. Wen, S.~Xia, and S.~Jin, ``Joint localization
  and environment sensing by harnessing {NLOS} components in {RIS}-aided
  mm{W}ave communication systems,'' \emph{IEEE Trans. Wireless Commun.},
  vol.~22, no.~12, pp. 8797--8813, Dec. 2023.

\bibitem{deban2009deterministic}
R.~Deban, H.~Boutayeb, K.~Wu, and J.~Conan, ``Deterministic approach for
  spatial diversity analysis of radar systems using near-field radar cross
  section of a metallic plate,'' \emph{IEEE Trans. Antennas Propag.}, vol.~58,
  no.~3, pp. 908--916, Mar. 2010.

\bibitem{wu2021through}
Q.~Wu, Z.~Lai, and M.~G. Amin, ``Through-the-wall radar imaging based on
  {B}ayesian compressive sensing exploiting multipath and target structure,''
  \emph{IEEE Trans. Comput. Imaging}, vol.~7, pp. 422--435, Apr. 2021.

\bibitem{stojanovic2008joint}
I.~Stojanovic, M.~Cetin, and W.~C. Karl, ``Joint space aspect reconstruction of
  wide-angle {SAR} exploiting sparsity,'' in \emph{Proc. SPIE Algorithms Synth.
  Aperture Radar Imagery XV}, vol. 6970, Mar. 2008, pp. 37--48.

\bibitem{ziniel2013dynamic}
J.~Ziniel and P.~Schniter, ``Dynamic compressive sensing of time-varying
  signals via approximate message passing,'' \emph{IEEE Trans. Signal
  Process.}, vol.~61, no.~21, pp. 5270--5284, Nov. 2013.

\bibitem{ziniel2012generalized}
J.~Ziniel, S.~Rangan, and P.~Schniter, ``A generalized framework for learning
  and recovery of structured sparse signals,'' in \emph{Proc. IEEE Stat. Signal
  Process. Workshop (SSP)}, Aug. 2012, pp. 325--328.

\bibitem{mahfouz2008investigation}
M.~R. Mahfouz, C.~Zhang, B.~C. Merkl, M.~J. Kuhn, and A.~E. Fathy,
  ``Investigation of high-accuracy indoor 3-{D} positioning using {UWB}
  technology,'' \emph{IEEE Trans. Microwave Theory Tech.}, vol.~56, no.~6, pp.
  1316--1330, Jun. 2008.

\bibitem{el2018high}
M.~El-Absi, A.~A. Abbas, A.~Abuelhaija, F.~Zheng, K.~Solbach, and T.~Kaiser,
  ``High-accuracy indoor localization based on chipless {RFID} systems at {TH}z
  band,'' \emph{IEEE access}, vol.~6, pp. 54\,355--54\,368, Sep. 2018.

\bibitem{huang2021reconf}
X.~Wang, Y.~Huang, J.~Yang, Y.~Han, and S.~Jin, ``Reconfigurable intelligent
  surface aided integrated communication and localization with a single access
  point,'' \emph{China Commun.}, [Online]. Available:
  http://www.cic-chinacommunications.cn, Apr. 2023.

\bibitem{zhang20223d}
Y.~Zhang, Z.~Zhang, X.~Tong, and C.~Huang, ``3{D} environment sensing with
  channel state information based on computational imaging,'' in \emph{Proc.
  IEEE ICC Workshops}, May 2022, pp. 842--847.

\bibitem{born2013principles}
M.~Born and E.~Wolf, \emph{Principles of Optics: Electromagnetic Theory of
  Propagation, Interference and Diffraction of Light}, 7th~ed.\hskip 1em plus
  0.5em minus 0.4em\relax Cambridge, U.K.: Cambridge Univ. Press, 1999.

\bibitem{Merrill1990radar}
M.~I. Skolnik, \emph{Radar Handbook}, 2nd~ed.\hskip 1em plus 0.5em minus
  0.4em\relax New York, USA: McGraw-Hill, 1990.

\bibitem{balanis2012advanced}
C.~A. Balanis, \emph{Advanced Engineering Electromagnetics}, 2nd~ed.\hskip 1em
  plus 0.5em minus 0.4em\relax Hoboken, NJ, USA: Wiley, 2012.

\bibitem{bellili2019generalized}
F.~Bellili, F.~Sohrabi, and W.~Yu, ``Generalized approximate message passing
  for massive {MIMO} mm{W}ave channel estimation with {L}aplacian prior,''
  \emph{IEEE Trans. Commun.}, vol.~67, no.~5, pp. 3205--3219, May 2019.

\bibitem{dempster1977maximum}
A.~P. Dempster, N.~M. Laird, and D.~B. Rubin, ``Maximum likelihood from
  incomplete data via the {EM} algorithm,'' \emph{J. Roy. Stat. Soc. B,
  Methodol.}, vol.~39, no.~1, pp. 1--22, 1977.

\end{thebibliography}

\end{document}